\documentclass[twocolumn]{aastex631}
\makeatletter
\@ifundefined{tablenum}{}{\let\tablenum\relax}
\makeatother
\usepackage{siunitx}
\usepackage{amsmath}
\usepackage{amssymb}
\usepackage{physics}
\usepackage{colortbl}
\usepackage{placeins}
\usepackage{bm}

\newcommand{\half}{\frac{1}{2}}
\newcommand{\pn}[1]{\left(#1\right)}

\shorttitle{Terrestrial Atmospheres Across Dynamical Regimes}
\shortauthors{Wirth, Powell, and Wordsworth}

\begin{document}

\title{Analytic Modeling of Tidally Locked Rocky Planet Atmospheres Across Dynamical Regimes}

\correspondingauthor{Christopher P. Wirth}
\email{cwirth@uchicago.edu}

\author[0000-0003-1656-011X]{Christopher P. Wirth}
\affiliation{Department of Astronomy \& Astrophysics, University of Chicago, 5640 S.~Ellis Avenue, Chicago, IL 60637, USA}

\author[0000-0002-4250-0957]{Diana Powell}
\affiliation{Department of Astronomy \& Astrophysics, University of Chicago, 5640 S.~Ellis Avenue, Chicago, IL 60637, USA}

\author[0000-0003-1127-8334]{Robin Wordsworth}
\affiliation{Harvard Paulson School of Engineering and Applied Sciences, 29 Oxford Street, Cambridge, MA 02138, USA}
\affiliation{Department of Earth and Planetary Sciences, Harvard University, 20 Oxford Street, Cambridge, MA 02138, USA}



\begin{abstract}

We present a new first-principles analytic approach to interpreting eclipses and phase curves of rocky planets. Observations with JWST have reported nondetections of atmospheres around the majority of hot rocky planets orbiting M dwarfs.
However, most of these “bare rock” inferences are based on models that are ill-suited to many currently observable planets, as they were developed for use on cooler, slower-rotating bodies. In particular, these models rely on the weak temperature gradient assumption, in which rotation is neglected and temperature gradients can be simply related to wind speeds. We find that this assumption may not be valid for over 40\% of terrestrials observed with JWST, including TRAPPIST-1b, GJ 367b, and TOI-2445b. Our simple new four-box model does not rely on this assumption, and instead allows the heat transport efficiency to be specified or follow scalings derived herein. This method is fast, interpretable, physically motivated, reproduces previous general circulation model results, and can be used as a starting point for more detailed modeling.
We observe that the longitudinal temperature structure of tidally locked terrestrials depends strongly on the atmospheric circulation. Considering the applicable range of atmospheric dynamical regimes, we find that a given planet’s nightside temperature can plausibly vary by 100s of Kelvin (from detectable to undetectable). Furthermore, a planet's dayside energy balance can display complex behavior, with degeneracies between surface pressure and dayside temperature. Illustrating an application to observations, we find that assumptions about atmospheric dynamics and longitudinal temperature structure can bias atmospheric constraints at the order-of-magnitude level.

\end{abstract}



\section{Introduction} \label{sec:intro}
 Within the last 15 years since the \textit{Kepler} mission \citep{borucki_kepler_2010} discovered its first rocky exoplanet, a substantial population of these planets has been identified, the study of which has supported significant theoretical work in interiors \citep[e.g.][]{zeng_mass-radius_2016}, orbital dynamics \citep[e.g.][]{dawson_correlations_2016}, evolutionary processes \citep[e.g.][]{ginzburg_core-powered_2018}, and atmospheres \citep[e.g.][]{showman_equatorial_2011}. The launch of the James Webb Space Telescope has enabled a new era of detailed system-level characterization for terrestrial planets. Eclipse depths and phase curves have been measured with sub-10 ppm precision, depending on stellar properties \citep[e.g.][]{weiner_mansfield_no_2024} allowing for the detection of thermal emission from over a dozen terrestrials. 

This emission is the most detectable for hotter planets contrasted with lower-luminosity stars, and thus the most robust measurements of rocky planet emission have been for those in close orbits around M dwarf stars.  However, M dwarfs have a long pre-main-sequence stage, and continue to flare and emit high-energy radiation throughout their lifetimes more frequently than their more massive counterparts \citep{gunther_stellar_2020}. This could potentially strip the atmospheres of planets around these stars \citep[e.g.][]{do_amaral_contribution_2022,van_looveren_airy_2024}. The mechanics of atmospheric escape are not fully understood, however, and it is possible that atmospheres could survive these harsh conditions \citep{nakayama_survival_2022} or be replenished by mantle outgassing \citep{katyal_effect_2020} or meteorite impacts \citep{kral_cometary_2018}. 

Taking advantage of the capabilities of JWST and other space-based observatories, recent work has aimed to tackle this issue from an empirical perspective: do we observe atmospheres on planets around M dwarfs? The answer thus far has appeared to be a resounding no. For example, \cite{kreidberg_absence_2019} reported an absence of a thick atmosphere on LHS 3844b using a thermal phase curve, \cite{greene_thermal_2023,zieba_no_2023} found similar results for the innermost planets of the TRAPPIST-1 system, and subsequent ``bare rock" detections around low-mass stars \citep[e.g.][]{xue_jwst_2024,luque_dark_2024,zhang_gj_2024} have continued to show evidence against the presence of substantive atmospheres on these planets. These observations have often been placed in the context of the ``cosmic shoreline," in which the survival of an atmosphere is dependent on the planet's escape velocity and cumulative high-energy instellation \citep{zahnle_cosmic_2017}, taking into account the latter axis' dependence on stellar type.


These conclusions are usually drawn from thermodynamic arguments, where large day-night temperature contrasts are consistent with those expected for airless bodies. If there were a substantive atmosphere, it would have redistributed heat around the planet---the thicker the atmosphere, the closer the planet gets to a uniform equilibrium temperature. This is connected to a major assumption made in previous work: that planets are in the Weak Temperature Gradient (WTG) regime, in which the planet's rotation is negligible compared to the atmospheric circulation, such that pressure and temperature gradients can be simply related to wind speeds.

Determining the atmospheric pressures and compositions that could result in observed temperatures requires two ingredients: a model that takes in possible atmospheres and produces outputs that can be compared with the data, and the evaluation of this model over a broad range of possible atmospheric configurations. Several different modeling approaches have been developed or adapted to address this problem.

General circulation models (GCMs) are the most comprehensive models of exoplanetary atmospheres, which vary in complexity and can include the interplay between atmospheric dynamics, radiation, chemistry, and aerosols. Their versatility allows them to be used to test underlying assumptions in simpler modeling frameworks \citep[e.g.][]{hammond_rotational_2021} or to resolve complex phenomena inaccessible to more streamlined models \citep[e.g.][]{sergeev_atmospheric_2020}. However, they are computationally intensive such that running a large grid of GCMs (particularly those that include the majority of relevant atmospheric phenomena) is currently infeasible for the purpose of detailed interpretation of observations, particularly as the number of planets observed with JWST increases. Additionally, sufficiently complicated GCMs suffer from a lack of interpretability such that simplified models are needed for detailed physical understanding.

Computationally efficient parameterized models have been developed for use in Bayesian inference retrieval frameworks, with the aim of enabling statistical conclusions from the data, such as \texttt{POSEIDON} \citep{macdonald_hd_2017,macdonald_poseidon_2023}, \texttt{PLATON} \citep{zhang_forward_2019, zhang_platon_2020,zhang_retrievals_2025}, and \texttt{TauREx} \citep{al-refaie_taurex_2021}. These simplified models are generally one-dimensional radiative-convective models that can be coupled to chemistry or aerosol modules, and have been widely applied to interpret observations \citep[e.g.][]{xue_jwst_2024,piaulet-ghorayeb_strict_2025,taylor_jwst_2025}. For the more uncertain data available for terrestrial planets, it is often the case that a variety of atmospheric or bare surface conditions modeled thus could reasonably explain the observations \citep[e.g.][]{lustig-yaeger_jwst_2023,alderson_jwst_2024}. In this case, it is important to ensure that models are not missing additional physical regions of phase space that could also well-explain the data and alter our interpretations.

The most intuitive entries on the spectrum of exoplanet atmospheric models are the order-of-magnitude or semi-analytic models, which aim to distill the essential physics of a system and describe regimes, scalings, and broad-scale behaviors. These models are physically motivated (while a free retrieval could yield seemingly non-physical solutions), intuitive, and fast. However, it is nontrivial to select the most important physics to include to make an analytic model feasible, so simplifying assumptions have to be made. A key challenge is ensuring that the assumptions made are appropriate for the regimes considered. Notably, the global WTG assumption is key in commonly used models \citep[e.g.][]{wordsworth_atmospheric_2015,koll_scaling_2022}, used as a starting point from which atmospheric behavior is derived. This may not be a universally safe assumption for terrestrial planets across all relevant regimes, as the short rotation periods experienced by many of the best targets observable with JWST increase the importance of rotation in their atmospheric dynamics.

In this paper, we examine the broader phase space of potential terrestrial atmospheres accessible with JWST by creating a flexible, physically motivated analytic model of atmospheric heat transport. In Section \ref{sec:WTG} we consider the WTG assumption and its applicability to terrestrial planets observed by JWST. Finding sufficient reason to explore the relaxation of this assumption, we present in Section \ref{sec:model} a new order-of-magnitude analytic model designed to be flexible over a broad range of dynamical regimes and atmospheric compositions. In Section \ref{sec:results}, we explore the results of calculations using this model and its application to a few commonly studied planets. In Section \ref{sec:discussion} we compare our results to other models, demonstrate their potential application to observations, and discuss limitations, possible improvements, and extensions of the model. We summarize our results in Section \ref{sec:conclusions}.

\section{The Weak Temperature Gradient Assumption} \label{sec:WTG}
The WTG assumption is one of the most common assumptions for nearly all semi-analytic models of atmospheres for a wide variety of planetary types \citep{pierrehumbert_palette_2011,perez-becker_atmospheric_2013,yang_low-order_2014,wordsworth_atmospheric_2015,koll_temperature_2016,zhang_effects_2017,komacek_atmospheric_2019,koll_scaling_2022}. This is because the WTG assumption vastly simplifies the required calculations, and many atmospheres with which we are most familiar operate roughly within this regime, such as the tropical regions of the Earth \citep{sobel_modeling_2000}.

However, even within the small sample size of bodies in our solar system, there exists a broad range of atmospheric dynamical regimes, which have first-order dependence on the planetary rotation period. Venus and Titan exhibit little horizontal temperature variation (although Venus' upper atmosphere does exhibit some) \citep{seiff_models_1985}, while Jupiter and Saturn display prominent zonal temperature bands \citep{orton_saturns_2005,odonoghue_global_2021}. On Earth, a mix of these regimes exists, where temperatures in the tropics are much more homogeneous than those near the poles. These differences can be ascribed to the different rotational regimes of these bodies; Venus and Titan complete a rotation on the order of tens to hundreds of days \citep{goldstein_rotation_1963,stiles_determining_2008}, Earth in one day, and Jupiter and Saturn in ten hours or less \citep{archinal_report_2011,desch_voyager_1981}.

Differences in rotation period are important in driving dynamics, as spatial variations in atmospheric heating lead to pressure gradients, which generally induce fluid flow to eliminate these gradients. In the ``slow-rotator" regime, this flow will efficiently transport mass and reduce horizontal temperature differences throughout the planet \citep{vallis_atmospheric_2017}, giving the assumption its name \citep{sobel_weak_2001}. Even for a planet with a strong heating contrast, this assumption allows for a simple relation between the forcing of the atmosphere and its response \citep{perez-becker_atmospheric_2013}. The assumption that tide-locked exoplanets are globally in WTG balance is ubiquitous in previous analytic studies, which often focused on habitable-zone planets with rotation periods of multiple days, and supported by dedicated GCM results such as \citet{merlis_atmospheric_2010}.

However, if the atmosphere is unable to quickly redistribute energy, these assumptions are not applicable and strong horizontal temperature differences can be sustained. This can happen if frictional forces and nonlinearities in the momentum equation are strong, or if Coriolis forces are non-negligible \citep{pierrehumbert_atmospheric_2019}. The latter is the most-studied case for non-WTG behavior \citep[e.g.][]{charney_note_1963,sobel_weak_2001,pierrehumbert_dynamics_2016}. In this case, Coriolis forces will inhibit meridional flow, reducing the atmosphere's ability to eliminate gradients, and potentially leading to equatorial jets such as those predicted and observed on hot Jupiters \citep{showman_atmospheric_2002,knutson_map_2007}.  

Planets most amenable to characterization with JWST tend to be on short-period orbits, and assuming tidal synchronization \citep{goldreich_q_1966,barnes_tidal_2017,pierrehumbert_atmospheric_2019,lyu_super-earth_2024} implies they also have short rotation periods. Thus, we investigate the rotational regime represented by the observed planet sample with a scale argument following \cite{pierrehumbert_atmospheric_2019}, which is similar to that of \citep{leconte_3d_2013}. That is, we compare the Rossby radius of deformation $L_d=c_0/\Omega,$ which can be thought of as the distance that a gravity wave can travel before being damped out by Coriolis forces, to the planet radius $r_p$. Here $\Omega$ is the angular frequency of rotation, and $c_0$ is the wave speed, which can be taken as an upper bound to be the isothermal speed of sound $c_0=\sqrt{RT},$ where $R$ is the specific gas constant and $T$ is atmospheric temperature. 

We define $\Lambda=L_d/r_p$ as the ratio of these distances, such that WTG behavior should be expected when $\Lambda\gg1$ and temperature gradients should persist when $\Lambda\ll1.$ Then, we have
\begin{equation}
    \Lambda=\sqrt{\frac{k_BT}{m}}\frac{P}{2\pi r_p},
\end{equation}
where $k_B$ is the Boltzmann constant and $P$ is the planet's orbital period (equal to the rotation period for a tidally locked body). We can then estimate $T\approx T_{eq},$ with
\begin{equation}
    T_{eq}=\left(\frac{
    L(1-A)}{16\sigma\pi a^2}\right)^{1/4}=T_{\star,\mathrm{eff}}\sqrt{\frac{R_\star}{2a}}(1-A)^{1/4},
\end{equation}
where $T_{eq}$ is the planet's equilibrium temperature, $R_\star$ is the stellar radius, $T_{\star,\mathrm{eff}}$ is the stellar effective temperature, $a$ is the orbital semi-major axis, $\sigma$ is the Stefan-Boltzmann constant, and $A$ is the planet's Bond albedo. If we also apply Kepler's Third Law $GM_\star P^2=4\pi^2 a^3,$ with $M_\star$ the stellar mass, we can write
\begin{equation}
    T_{eq}=T_{\star,\mathrm{eff}}\pn{\frac{\pi^2 R_\star^3}{2GM_\star P^2}}^{1/6}(1-A)^{1/4}
\end{equation}
and
\begin{equation}
    \Lambda=\sqrt{\frac{k_B T_{\star,\mathrm{eff}}}{m}}\pn{\frac{\pi^2 R_\star^3}{2GM_\star P^2}}^{1/12}\pn{\frac{P}{2\pi r_p}}(1-A)^{1/8}
\end{equation}
\begin{multline}
    \Lambda=1.66\pn{\frac{T_{\star,\mathrm{eff}}}{\SI{5777}{K}}}^{1/2}\pn{\frac{m}{\SI{28}{amu}}}^{-1/2}\pn{\frac{R_\star}{R_\sun}}^{1/4}\\
    \pn{\frac{M_\star}{M_\sun}}^{-1/12}\pn{\frac{r_p}{R_\earth}}^{-1}\pn{\frac{P}{\SI{}{day}}}^{5/6}(1-A)^{1/8}.
\end{multline}
$\Lambda$ has the strongest dependence, therefore, on the planet's radius, as well as on its orbital period. The stellar type and mean molecular weight (MMW) of the atmosphere also contribute. 

\begin{figure}
    \centering
    \includegraphics[width=1.0\linewidth]{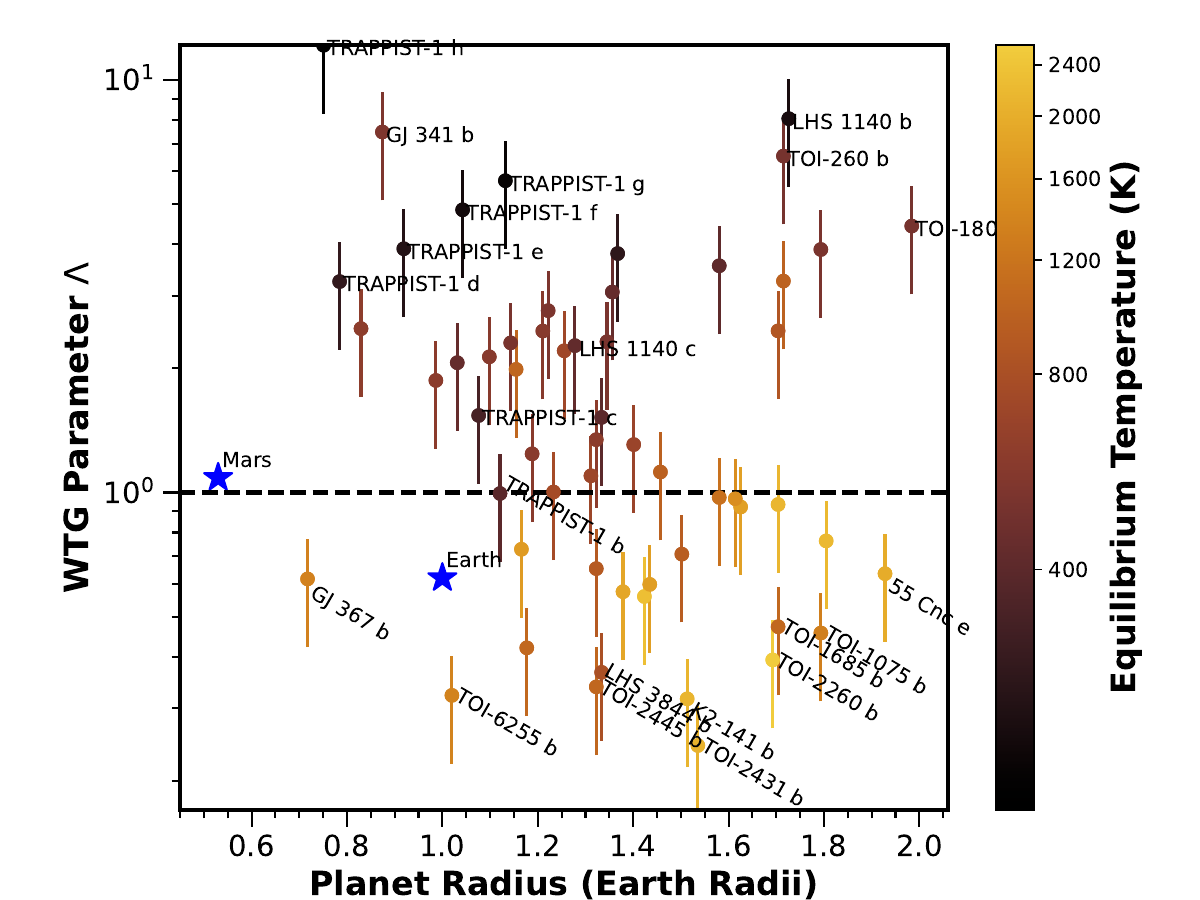}
    \caption{Estimated values of the WTG parameter $\Lambda$ for planets $r<2R_\earth$ observed with JWST. $\Lambda\ll1$ implies non-WTG behavior and non-negligible temperature gradients, while $\Lambda\gg1$ implies WTG behavior is likely dominant. The points correspond to an N$_2$ mean molecular weight, while the upper error bars correspond to H$_2$O and the lower to SiO$_2$. Points further from $\Lambda=1$ (or particularly well-known bodies) are labeled with the planet name. The colorbar represents the planet's equilibrium temperature, with zero albedo assumed. Blue starred points are Solar System bodies for comparison.}
    \label{fig:lambda_calc}
\end{figure}

As expected, the WTG assumption is the least valid for large, rapidly rotating planets with high-MMW atmospheres around cool stars. Many of the planets that are more easily observed in emission with JWST have those exact biases \citep{kempton_framework_2018}. We consider the planets with approved observations through Cycle 4 using TrExoLiSTS \citep{nikolov_trexolists_2022}, and obtain planet and system parameters using NASA's Exoplanet Archive \citep{nasa_exoplanet_archive_confirmed_2019,christiansen_nasa_2025}. Thus, it is simple to estimate $\Lambda$ for this sample, the results of which are shown in Figure \ref{fig:lambda_calc}. We conservatively assume a zero-albedo planet, which increases the assumed temperature and thus favors WTG behavior. We consider all planets with $r_p<2R_\earth$. Because $\Lambda$ depends on atmospheric mean molecular weight, which is unknown, we compute this value for three different potential compositions: H$_2$O (18 g/mol), N$_2$ (28 g/mol), and SiO$_2$ (60 g/mol). Hydrogen and helium-dominated atmospheres are often ruled out on terrestrial exoplanets from density and mass-loss considerations \citep[e.g.][]{brinkman_toi-561_2023}, but would generally trend more towards WTG behavior if present.

We observe that for a medium-weight N$_2$ atmosphere, 39\% of JWST-observed planets have $\Lambda < 1$, disfavoring global WTG behavior. This fraction decreases to 30\% for a lighter H$_2$O atmosphere and increases to 50\% if these planets have heavier SiO$_2$ atmospheres. This is a biased sample as these planets are easier to observe, but it demonstrates that these observability considerations lead to observing planets with short orbital periods, for which underlying assumptions of global WTG balance need to be revisited.

\section{Analytic Model} \label{sec:model}
Having motivated the relaxation of the WTG assumption, we turn now to the development of an analytic framework for evaluating heat redistribution on a tidally locked rocky planet without using this assumption. Our approach is based on \citet{wordsworth_atmospheric_2015}, which we refer to as W15 for brevity, which in turn built on the analyses of \citet{pierrehumbert_palette_2011} and \citet{yang_low-order_2014}. We note that \citet{joshi_simulations_1997} performed an analysis of tidally locked terrestrial planet heat redistribution before any extrasolar rocky planets had been discovered. Additionally, \citet{koll_deciphering_2015} presented a compelling reduction of the problem to key dimensionless parameters, which also motivated this work as an exploration of the hot and rapidly rotating regime which was a major caveat to their results. 

\begin{figure}
    \centering
    \includegraphics[width=0.9\linewidth]{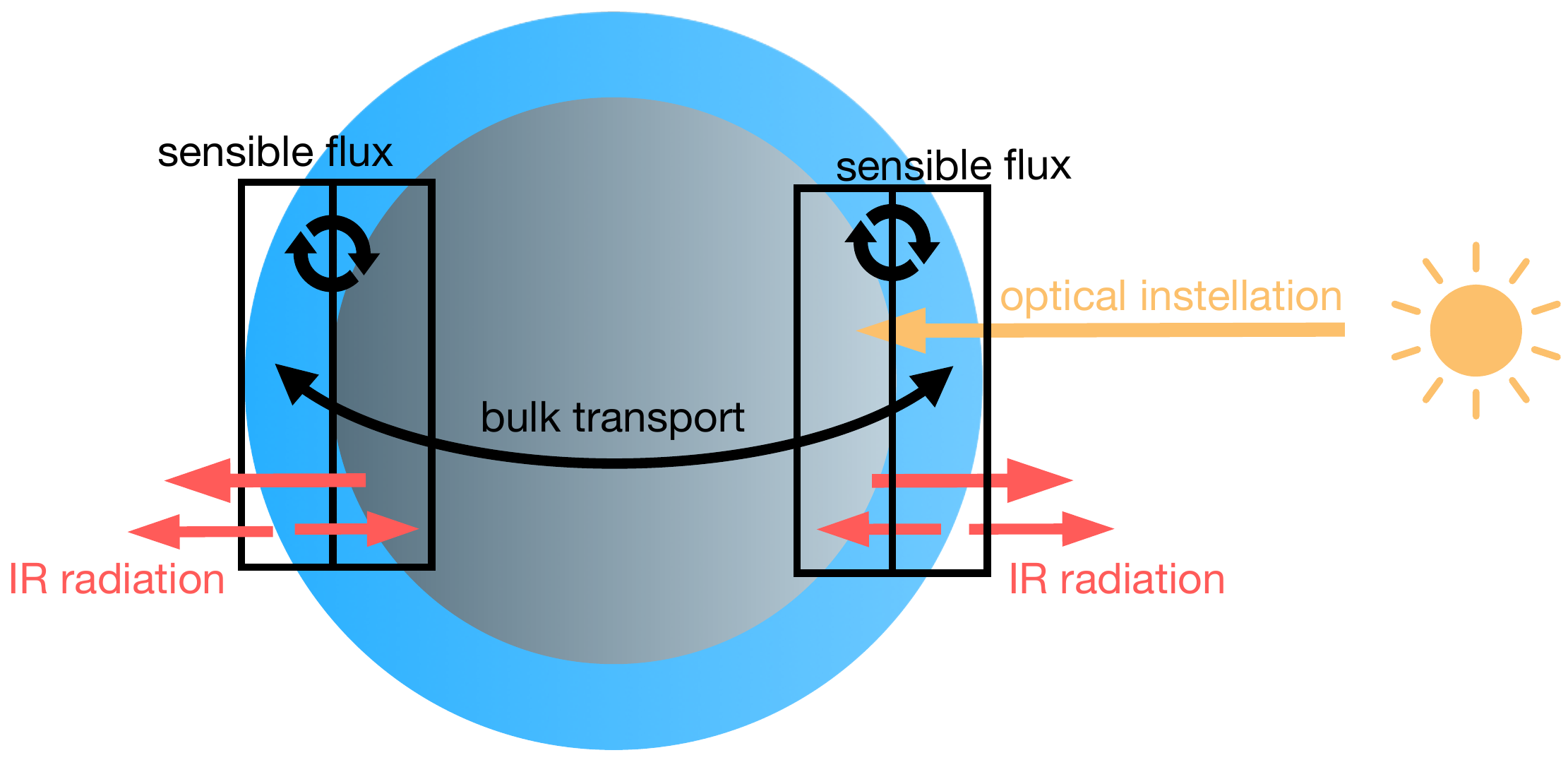}
    \caption{Schematic diagram of the dominant mechanisms of heat transport on a tidally locked terrestrial planet in this model: stellar radiation, atmospheric bulk fluid transport, surface heating, and re-radiation.}
    \label{fig:cartoon}
\end{figure}

The most significant change in our model is that we allow temperature to vary longitudinally, thus allowing us to describe atmospheres outside of the WTG regime without a single homogeneous atmosphere temperature. Given the stability of the tidally locked regime, we apply a steady-state four-box model of the atmosphere-surface system. Figure \ref{fig:cartoon} gives a schematic of the model, where ``boxes'' are drawn to represent the dayside and nightside hemispheres at the surface and in the atmosphere, considered as a whole. Our goal is to derive the temperatures in these regions as a function of planetary and atmospheric parameters.

We assume the atmosphere is transparent in the visible band, and can be described by a single gray optical depth in the infrared band. We also assume that the atmospheric mass is the same between the day and night sides, i.e. we have a single surface pressure. Because atmospheric escape is expected to be important for the close-in planets where the WTG assumption is least valid, and observations so far have not shown evidence for thick atmospheres, we focus on the optically thin regime. Without the WTG assumption prescribing day-to-night advection, a major assumption we make is to instead relate the heat transport speed to the sound speed by a parameter $\alpha,$ such that $U\sim\alpha c_s$. We explore the possible parameter space of $\alpha$ as well as possible scaling relations for its behavior.

\subsection{Steady-state equations}\label{sec:steadystate}
Because the exoplanetary systems we observe are billions of years old, and tidal circularization timescales are comparatively short, we expect that the thermal structure of the atmosphere should remain broadly consistent over time, and thus we should use a model of the system that reflects this. Thus, we begin by describing the steady-state energy balance in a model atmosphere.

The local surface energy balance can be written as:
\begin{equation}
    \sigma T_s^4=(1-A)F\cos\theta_z+\mathrm{GLR}+C_D c_p \rho\abs{\mathbf{w}}(T_a-T_s),
    \label{eq:localsurf}
\end{equation}
where $\sigma$ is the Stefan-Boltzmann constant, $T_s$ is the local surface temperature, $T_a$ is the local atmospheric temperature, $A$ is the planetary albedo, $F$ is the stellar flux, $\theta_z$ is the stellar zenith angle, GLR is the infrared flux received from the atmosphere, $C_D$ is the surface drag coefficient, $c_p$ is the atmospheric heat capacity at constant pressure, $\rho$ is the atmospheric density, and $\abs{\mathbf{w}}$ is the wind speed at the surface. Here, the left side of the equation represents the energy emitted from the surface as blackbody radiation, the first term on the right side represents the energy absorbed from stellar insolation, and the second term is re-radiation down from the atmosphere. The last term in equation (\ref{eq:localsurf}) represents the sensible heat flux between surface and atmosphere through drag from the surface on the atmosphere, using the expression from W15. The rate of energy transfer via this mechanism increases when there is more drag, when the atmosphere is more dense or can hold more heat, when surface wind speeds are faster, and when the temperature difference is larger.

As for the local atmospheric steady-state, we have:
\begin{equation}
    \mathrm{OAR}+\mathrm{GLR}=\mathcal{A}\sigma T_s^4+C_Dc_p\rho\abs{\mathbf{w}}(T_s-T_a)-\frac{p_s}{r_pg}\abs{\mathbf{v}}k_Tc_p(\Delta T_a),
\end{equation}
where OAR is the infrared emission lost to space from the atmosphere, $\mathcal{A}$ is the fraction of the ground emission absorbed by the atmosphere, $p_s$ is the surface pressure, $r_p$ is the planet radius, $g$ is the surface gravity, $\abs{\mathbf{v}}$ is the speed of bulk atmospheric flow, $k_T$ is a turbulent diffusivity coefficient, and $\Delta T_a$ is the local atmospheric temperature difference in the direction of the bulk flow. Here, the left side of the equation represents the energy lost by thermal radiation of the atmosphere (upwards and downwards). The first term on the right side is the energy absorbed from the surface's thermal flux, the second term is the reverse of the last term in equation (\ref{eq:localsurf}), and the last term is the energy exchange from bulk motions in the atmosphere, following \citet{pierrehumbert_principles_2010}. Similar to the surface-atmosphere exchange term, this is more efficient if there is more atmosphere, the atmosphere can hold more heat, when wind speeds are faster, and when temperature gradients are steeper. We can apply the ideal gas law $\rho TR=p$ to write
\begin{equation}
    \mathrm{OAR}+\mathrm{GLR}=\mathcal{A}\sigma T_s^4+C_Dc_p\rho\abs{\mathbf{w}}(T_s-T_a)-\frac{\rho RT_a}{r_pg}\abs{\mathbf{v}}k_Tc_p(\Delta T_a).
    \label{eq:localatm}
\end{equation}

To reduce the dimensionality of this model to allow for simple calculations, and because most of the observational data for rocky planets is hemispherically averaged, we consider the integration of these equations over the full dayside and nightside hemispheres. We define mean hemispheric surface and atmosphere emissions:
\begin{equation}
    \begin{split}
    B_{s,d}=\frac{\int_d\sigma T_s^4\dd{A}}{2\pi r_p^2},\quad B_{s,n}=\frac{\int_n\sigma T_s^4\dd{A}}{2\pi r_p^2},\\
    B_{a,d}=\frac{\int_d\sigma T_a^4\dd{A}}{2\pi r_p^2},\quad B_{a,n}=\frac{\int_n\sigma T_a^4\dd{A}}{2\pi r_p^2},
    \end{split}
\end{equation}
where $\int_d\dd{A}$ is a surface integral over the dayside of the planet, $\int_n\dd{A}$ is over the nightside, and $r_p$ is the planetary radius.

Integration of equation (\ref{eq:localsurf}) over the dayside and dividing through by $2\pi r_p^2$ yields
\begin{equation}
    B_{s,d}=\frac{1}{2}(1-A)F+\left(\overline{\mathrm{GLR}}\right)_d+C_Dc_p\rho_d\overline{\abs{\mathbf{w}}(T_{a,d}-T_{s,d})},
    \label{eq:integ_day_surf}
\end{equation}
where the overline represents a dayside hemispheric average. We assume a constant atmospheric density, drag coefficient, and heat capacity over the hemisphere, but allow for variations in temperature and surface wind speed. Performing the same operation on equation (\ref{eq:localsurf}) over the nightside yields
\begin{equation}
    B_{s,n}=\left(\overline{\mathrm{GLR}}\right)_n+C_D c_p \rho_n \overline{\abs{\mathbf{w}}(T_{a,n}-T_{s,n})},
    \label{eq:integ_night_surf}
\end{equation}
where here the overlines represent nightside hemispheric averages.

Similarly, we integrate \eqref{eq:localatm} over the dayside and nightside to get
\begin{multline}
    \left(\overline{\mathrm{OAR}}\right)_d+\left(\overline{\mathrm{GLR}}\right)_d=\mathcal{A}B_{s,d}-\frac{R}{r_pg}c_p\rho_dk_T\overline{\abs{\mathbf{v}}T_{a,d}(\Delta T_a)}\\+C_Dc_p\rho_d\overline{\abs{\mathbf{w}}(T_{s,d}-T_{a,d})}
    \label{eq:integ_day_atm}
\end{multline}
\begin{multline}
\left(\overline{\mathrm{OAR}}\right)_n+\left(\overline{\mathrm{GLR}}\right)_n=\mathcal{A}B_{s,n}+\frac{R}{r_pg}\rho_dk_T\overline{\abs{\mathbf{v}}T_{a,d}(\Delta T_a)}\\+C_Dc_p\rho_n\overline{\abs{\mathbf{w}}(T_{s,n}-T_{a,n})}
    \label{eq:integ_night_atm}
\end{multline}
The $\rho_d$ in equation (\ref{eq:integ_night_atm}) is present on the nightside because energy is transferred from the dayside via advection, and thus the amount of heat scales with the dayside atmospheric density.

We now assume that the atmosphere is optically thin in the infrared, with optical depth given by $\tau$. In this case, $\mathcal{A}\approx\tau,$ and $\mathrm{OAR}\approx\mathrm{GLR}\approx\tau B_a.$ We make the further assumption that each hemisphere of the atmosphere can be approximated by a single mean temperature, allowing us to write equations (\ref{eq:integ_day_surf}) and (\ref{eq:integ_night_surf}) as 
\begin{equation}
    B_{s,d}=\frac{1}{2}(1-A)F+\tau B_{a,d}+C_Dc_p\rho_d\overline{\abs{\mathbf{w}}(T_{a,d}-T_{s,d})}
    \label{eq:thin_surf_day}
\end{equation}
\begin{equation}
    B_{s,n}=\tau B_{a,d}+C_Dc_p\rho_n\overline{\abs{\mathbf{w}}(T_{a,n}-T_{s,n})}.
    \label{eq:thin_surf_night}
\end{equation}
Dividing through by $\tau,$ equations (\ref{eq:integ_day_atm}) and (\ref{eq:integ_night_atm}) become
\begin{multline}
    2B_{a,d}=B_{s,d}+\frac{c_p\rho_d}{\tau}\left(-\frac{Rk_T}{r_pg}\overline{\abs{\mathbf{v}}T_{a,d}(\Delta T_a)}\right.\\\left.+C_D\overline{\abs{\mathbf{w}}(T_{s,d}-T_{a,d})}\right)
\end{multline}

\begin{multline}
    2B_{a,n}=B_{s,n}+\frac{c_p\rho_d}{\tau}\left(\frac{Rk_T}{r_pg}\overline{\abs{\mathbf{v}}T_{a,d}(\Delta T_a)}\right.\\\left.+C_D\frac{\rho_n}{\rho_d}\overline{\abs{\mathbf{w}}(T_{s,n}-T_{a,n})}\right).
\end{multline}
To capture potential weighting or geometric factors in the hemispheric averages, we define dimensionless factors $\xi$ and $\chi$ such that we can rewrite $\overline{\abs{\mathbf{v}}T_{a,d}(\Delta T_a)}=\xi\overline{\abs{\mathbf{v}}}\overline{T_{a,d}}
\left(\overline{T_{a,d}}-\overline{T_{a,n}}\right)$ and we can rewrite $    \overline{\abs{\mathbf{w}}(T_{a,d}-T_{s,d})}=\chi\overline{\abs{\mathbf{w}}}\left(\overline{T_{a,d}}-\overline{T_{s,d}}\right)$.
We define $\xi$ thus because we expect that last term acts as a telescoping series such that intermediate values of $T_a$ cancel, and only the boundary values contribute. Since we only integrate over a hemisphere, the outer boundary of integration has an atmospheric temperature between that of the day and night side, and so the integrated term should have an extra numeric factor. These are estimated in Appendix \ref{sec:app_xi}.

For a hemispherically averaged optical depth, we take
\begin{equation}
    \tau=\frac{2\kappa p}{g},
    \label{eq:tau}
\end{equation}
where $\kappa$ is the gas opacity, $p$ is the surface pressure, and $g$ is the surface gravity. This scales the amount of infrared absorption by the gas opacity, as well as by the total mass of atmosphere $p/g$. In particular, we have a linear dependence on $p$ because for thin atmospheres we might expect an equivalent gray gas opacity to be dominated by individual absorption lines \citep{pierrehumbert_principles_2010}. This is not necessarily always a good approximation, such as for CO above about 0.1 bar, where the gray opacity may be dominated by the overall band shape \citep{koll_earths_2018,koll_scaling_2022}.

Then, we pull out a factor of $\overline{T_{a,d}}$ from the last term of equation (\ref{eq:thin_surf_day}), using the ideal gas law $\rho TR=p$ with $R$ the specific gas constant and equation (\ref{eq:tau}) to write
\begin{equation}
    B_{s,d}=\half(1-A)F+\tau B_{a,d}+\tau\frac{\chi C_D c_p g}{2\kappa R}\abs{\mathbf{w}}(1-T_{s,d}/T_{a,d}).
    \label{eq:ig_surf_day}
\end{equation}
From now on, temperature and wind variables signify hemispheric averages. The same process can be applied to equation (\ref{eq:thin_surf_night}), where we divide through by $\tau$ and pull out a factor of $T_{a,n}$ to get
\begin{equation}
    \frac{B_{s,n}}{\tau}=B_{a,n}+\frac{\chi C_Dc_pg}{2\kappa R}\abs{\mathbf{w}}\left(1-T_{s,n}/T_{a,n}\right).
    \label{eq:ig_surf_night}
\end{equation}
In the same way, we also have 
\begin{multline}
    2B_{a,d}=B_{s,d}+\frac{c_pg}{2\kappa R}\left(\xi\frac{RT_{a,d}k_T}{r_pg}\abs{\mathbf{v}}(T_{a,n}/T_{a,d}-1)\right.\\\left.+\chi C_D\abs{\mathbf{w}}(T_{s,d}/T_{a,d}-1)\right)
    \label{eq:ig_atm_day}
\end{multline}
\begin{multline}
    2B_{a,n}=B_{s,n}+\frac{c_pg}{2\kappa R}\left(\xi\frac{RT_{a,d}k_T}{r_pg}\abs{\mathbf{v}}(1-T_{a,n}/T_{a,d})\right.\\\left.+\chi C_D\abs{\mathbf{w}}(T_{s,n}/T_{a,n}-1)\right),
    \label{eq:ig_atm_night}
\end{multline}
where the atmospheric density ratio is canceled in the last term due to our assumption of a uniform surface pressure and the factor of $T_{a,n}/T_{a,d}$.

We also define a ``maximal dayside surface temperature" $T_0=T_{\mathrm{eff},\star}(\frac{1-A}{2})^{1/4}\left(\frac{a}{R_\star}\right)^{-1/2}$ to represent the mean temperature over the dayside surface if there were no atmosphere, with $B_0=\sigma T_0^4$.
If we then define the dimensionless temperature ratios 
\begin{equation}
    \tilde{T}_0=\frac{T_{s,d}}{T_0},\quad\tilde{T}_1=\frac{T_{a,d}}{T_{s,d}},\quad\tilde{T}_2=\frac{T_{a,n}}{T_{a,d}},\quad\tilde{T}_3=\frac{T_{s,n}}{T_{a,n}},
    \label{eq:T_defns}
\end{equation}
we can rewrite equations (\ref{eq:ig_surf_day}-\ref{eq:ig_atm_night}) as
\begin{equation}
    B_{s,d}=B_0+\tau B_{a,d}+\tau\frac{\chi C_Dc_pg}{2\kappa R}\abs{\mathbf{w}}(1-\tilde{T}_1^{-1})
    \label{eq:temp_surf_day}
\end{equation}
\begin{equation}
    \frac{B_{s,n}}{\tau}=B_{a,n}+\frac{\chi C_Dc_pg}{2\kappa R}\abs{\mathbf{w}}(1-\tilde{T}_3)
    \label{eq:temp_surf_night}
\end{equation}
\begin{multline}
    2B_{a,d}=B_{s,d}+\frac{\chi c_pg}{2\kappa R}\left(\frac{\xi}{\chi}C_T\tilde{T}_0\tilde{T}_1\abs{\mathbf{v}}(\tilde{T}_2-1)\right.\\\left.+C_D\abs{\mathbf{w}}(\tilde{T}_1^{-1}-1)\right)
    \label{eq:temp_atm_day}
\end{multline}
\begin{multline}
    2B_{a,n}=B_{s,n}+\frac{\chi c_pg}{2\kappa R}\left(\frac{\xi}{\chi}C_T\tilde{T}_0\tilde{T}_1\abs{\mathbf{v}}(1-\tilde{T}_2)\right.\\\left.+C_D\abs{\mathbf{w}}(\tilde{T}_3-1)\right),
    \label{eq:temp_atm_night}
\end{multline}
where we define $C_T=RT_0k_T/r_p g$ for symmetry with $C_D$.

Due to our uncertainty in the mechanism and efficiency of the day/night transport, we quantify it by assuming that its wind speed, $\abs{\mathbf{v}}$, is proportional to the sound speed in the dayside atmosphere by some constant $\alpha$, so we can write 
\begin{equation}
    \abs{\mathbf{v}}=\alpha c_s=\alpha\sqrt{c_p R T_{a,d}/c_v}
\end{equation}
Defining a characteristic velocity
\begin{equation}
    W_0=\frac{2\kappa R B_0}{\chi C_D c_p g}=\frac{\kappa R (1-A)F}{\chi C_D c_p g},
\end{equation}

we can define a dimensionless velocity $\tilde{W}=\abs{\mathbf{w}}/W_0$, as well as 
\begin{equation}
\tilde{V}=\frac{\alpha\sqrt{c_p R T_{0}/c_v}}{W_0}\implies\frac{\abs{\mathbf{v}}}{W_0}=\tilde{V}\tilde{T}_0^{\half}\tilde{T}_1^{\half}.
\end{equation}
We collect drag and integration constants into
\begin{equation}
    \tilde{C}=\frac{\xi C_T}{\chi C_D}
\end{equation}
and are thus able to write equations (\ref{eq:temp_surf_day}-\ref{eq:temp_atm_night}) as fully non-dimensionalized equations:
\begin{equation}
    \tilde{T}_0^4-1=\tau\tilde{T}_0^4\tilde{T}_1^4+\tau\tilde{W}(1-\tilde{T}_1^{-1}
).    \label{eq:eq0}
\end{equation}
\begin{equation}
    \frac{\tilde{T}_3^4}{\tau}-1=\tilde{T}_2^{-4}\tilde{T}_1^{-4}\tilde{T}_0^{-4}\tilde{W}(1-\tilde{T}_3)
    \label{eq:eq1}
\end{equation}
\begin{equation}
    2\tilde{T}_1^4-1=\tilde{T}_0^{-4}\left(\tilde{C}\tilde{V}(\tilde{T}_0\tilde{T}_1)^{3/2}(\tilde{T}_2-1)+\tilde{W}(\tilde{T}_1^{-1}-1)\right)
    \label{eq:eq2}
\end{equation}
\begin{equation}
    2-\tilde{T}_3^{4}=\left(\tilde{T}_2\tilde{T}_1\tilde{T}_0\right)^{-4}\left(\tilde{C}\tilde{V}(\tilde{T}_0\tilde{T}_1)^{3/2}(1-\tilde{T}_2)+\tilde{W}(\tilde{T}_3-1)\right).
    \label{eq:eq3}
\end{equation}

\subsection{Thermodynamic Equation}\label{sec:thermo}
To close the system, as we currently have four equations and five unknowns ($\tilde{T}_0$, $\tilde{T}_1,$ $\tilde{T}_2$, $\tilde{T}_3$, and $\tilde{W}$), we use the thermodynamic equation, which can be written
\begin{equation}
    \frac{\mathrm{D}I}{\mathrm{D}t}+\frac{p}{\rho}\mathbf{\nabla}\cdot\mathbf{u}=\mathcal{H},
\end{equation}
where $I$ is internal energy, $\mathcal{H}$ is the net heating or cooling rate per unit mass of the atmosphere, $\mathbf{u}$ is the net wind velocity, and $\frac{\mathrm{D}}{\mathrm{D}t}$ is the advective derivative. We expand this derivative to write
\begin{equation}
    \pdv{I}{t}+u_x\pdv{I}{x}+u_y\pdv{I}{y}+u_z\pdv{I}{z}+\frac{p}{\rho}\mathbf{\nabla}\cdot\mathbf{u}=\mathcal{H}.
\end{equation}
Here, we choose a local coordinate system in the dayside atmosphere such that $x$ increases away from the substellar point, $y$ is perpendicular to $x$ along the surface, and $z$ represents altitude. Because our model is steady-state, $\pdv{I}{t}=0.$ We assume temperature varies mostly as a function of $x$ such that $\pdv{I}{y}$ is negligible, and since the atmosphere is optically thin we assume $\pdv{I}{z}$ is also negligible and that $\mathbf{u}$ has only a small component in the $z$-direction to write
\begin{equation}
    u_x c_v\pdv{T}{x}+RT_{a,d}\mathbf{\nabla}\cdot\mathbf{u}=\mathcal{H}.
    \label{eq:therm2}
\end{equation}
In finding this relation we also apply the ideal gas law and the result that $\dd{I}=c_v\dd{T}$ for an ideal gas, where $c_v$ is the heat capacity at constant volume. In our simplified model, gas flows from the dayside to the nightside with velocity $\mathbf{v}$ and returns from the nightside to the dayside with velocity $\mathbf{w}$ (see Figure 7 of W15 for an illustration). Thus, the component of the gas velocity in the $x$ direction is approximately $\abs{\mathbf{v}}-\abs{\mathbf{w}}.$ Similarly, over the dayside atmosphere the average derivative of temperature is approximately

\begin{equation}
    \pdv{T}{x}\approx\frac{1}{r_p}\pn{\frac{T_{a,d}+T_{a,n}}{2}-T_{a,d}}=\frac{1}{2r_p}(T_{a,n}-T_{a,d}).
\end{equation}

We also have that over the dayside atmosphere the average divergence is approximately $\mathbf{\nabla}\cdot\mathbf{u}\approx(\abs{\mathbf{v}}-\abs{\mathbf{w}})/r_p$.
Substituting these into  equation (\ref{eq:therm2}) and dividing through by $\frac{RT_{a,d}}{r_p}$ we have
\begin{equation}
    \frac{3}{2}\pn{\abs{\mathbf{v}}-\abs{\mathbf{w}}}\pn{\tilde{T}_2-1}+\pn{\abs{\mathbf{v}}-\abs{\mathbf{w}}}=\frac{r_p}{RT_{a,d}}\mathcal{H},
    \label{eq:therm_vel}
\end{equation}
using the approximation $c_v/R=(\gamma-1)^{-1}\approx3$ for a diatomic or triatomic ideal gas with $\gamma$ as the heat capacity ratio. We then explore $\mathcal{H},$ taking into account all the possible sources of heating or cooling to write
\begin{multline}
    \mathcal{H}=\kappa\sigma T_{s,d}^4-2\kappa\sigma T_{a,d}^4+\frac{g\xi C_T\tilde{T}_0\tilde{T}_1\abs{\mathbf{v}}c_p(T_{a,n}-T_{a,d})}{RT_{a,d}}\\+\frac{g\chi C_D\abs{\mathbf{w}}c_p(T_{s,d}-T_{a,d})}{RT_{a,d}},
\end{multline}
where we divide through by a factor $p/g$ to convert the rates from equation (\ref{eq:temp_atm_day}) into heating/cooling per unit mass. We divide equation (\ref{eq:therm_vel}) through by $W_0/2$, and examine the left side first:
\begin{multline}
    3\pn{\tilde{V}\tilde{T}_0^{\half}\tilde{T}_1^{\half}-\tilde{W}}\pn{\tilde{T}_2-1}+2\pn{\tilde{V}\tilde{T}_0^{\half}\tilde{T}_1^{\half}-\tilde{W}}\\=\pn{3\tilde{T}_2-1}\pn{\tilde{V}\tilde{T}_0^{\half}\tilde{T}_1^{\half}-\tilde{W}}.
\end{multline}
As for the right side, the first two terms become
\begin{multline}
    \frac{2r_p}{RT_{a,d}W_0}\pn{\kappa\sigma T_{s,d}^4-2\kappa\sigma T_{a,d}^4}=\\\frac{2\chi C_D c_p r_p g\sigma T_{s,d}^4}{R^2 T_{a,d}(1-A)F}\pn{1-2\tilde{T_1}^4}
    =\frac{\chi C_Dc_pr_pg}{R^2T_{a,d}}\tilde{T}_0^4\pn{1-2\tilde{T_1}^4}
\end{multline}
and the second two become
\begin{multline}
    \frac{2r_p}{RT_{a,d}W_0}\left(\frac{g\xi C_T\tilde{T}_0\tilde{T}_1\abs{\mathbf{v}}c_p(T_{a,n}-T_{a,d})}{RT_{a,d}}\right.\\\left.+\frac{g\chi C_D\abs{\mathbf{w}}c_p(T_{s,d}-T_{a,d})}{RT_{a,d}}\right)
    \\=\frac{2\chi C_Dc_p r_p g}{R^2T_{a,d}}\pn{\tilde{C}\tilde{V}(\tilde{T}_0\tilde{T}_1)^{3/2}\pn{\tilde{T}_2-1}+\tilde{W}\pn{\tilde{T}_1^{-1}-1}}.
\end{multline}
Finally, we define
\begin{equation}
    \tilde{L}=\frac{\chi C_D c_p r_p g}{R^2T_0}
\end{equation}
such that we can write
\begin{multline}
    \tilde{L}\tilde{T}_0^{-1}\tilde{T}_1^{-1}\left(\tilde{T}_0^4-2\tilde{T}_0^4\tilde{T}_1^4+2\tilde{C}\tilde{V}(\tilde{T}_0\tilde{T}_1)^{3/2}\pn{\tilde{T}_2-1}\right.\\\left.+2\tilde{W}\pn{\tilde{T}_1^{-1}-1}\right)\\=\pn{3\tilde{T}_2-1}\pn{\tilde{V}\tilde{T}_0^{\half}\tilde{T}_1^{\half}-\tilde{W}}.
    \label{eq:eq4}
\end{multline}

Together, equations (\ref{eq:eq0}), (\ref{eq:eq1}), (\ref{eq:eq2}), (\ref{eq:eq3}), and (\ref{eq:eq4}) form a closed system that can be solved for $\tilde{T}_0$, $\tilde{T}_1,$ $\tilde{T}_2$, $\tilde{T}_3,$ and $\tilde{W}$ by numerical methods. Given these values, we can evaluate $\abs{\mathbf{w}}=\tilde{W}W_0$ or
\begin{equation}
    T_{s,n}=\tilde{T}_3T_{a,n}=\tilde{T}_2\tilde{T}_3T_{a,d}=\tilde{T}_1\tilde{T}_2\tilde{T}_3T_{s,d}=\tilde{T}_0\tilde{T}_1\tilde{T}_2\tilde{T}_3T_0.
\end{equation}
This allows for simple calculation of the average temperatures of four different regions of the planet, as a function of the planet's maximal (or equilibrium) temperature. Thus, we have a system that can be solved to give a first-principles picture of the temperature structure of a planet with given atmospheric makeup. 

We solve this system using the ``Trust Region Reflective" least-squares algorithm \citep{branch_subspace_1999} implemented in \texttt{scipy}. This algorithm is well-suited for the low-rank bounded optimization problem, as we restrict our outputs to have $0\leq\tilde{T}_i\leq1,$ except for $\tilde{T}_0$ which we allow to be greater than 1, and $\tilde{W}\geq0$. 
We select an initial guess for the optimization by pre-computing a grid of solutions as a function of $\tilde{V}$ using Mathematica's robust NSolve function, and fitting a power law to the results for each variable. When using these starting points, we find each \texttt{scipy} optimization to take approximately 30 ms, which could be improved through a more robust initial guess selection or a more tailored optimization algorithm.

\subsection{Parameters and values} \label{sec:parameters}
To solve for these quantities and evaluate temperature contrasts, we must first know the dimensionless parameters $\tilde{L},$ $\tilde{C},$ $\tau,$ and $\tilde{V}$. While these are defined in \ref{sec:steadystate} and \ref{sec:thermo}, they depend on various atmospheric and planetary quantities. These are laid out in Table \ref{table:parameters} along with the fiducial values we adopt in this work. Here, we give more detail about these choices.

\begin{deluxetable*}{l | c c c c c}[t!]
\tabletypesize{\footnotesize}
\tablecaption{Model parameters and fiducial values \label{table:parameters}}
\tablehead{
\colhead{Parameter} & \colhead{Symbol} & \colhead{Fiducial Value} & \colhead{Unit} & \colhead{Source} & \colhead{Reference}
}

\startdata
Planet mass                      & $m_p$    & --                   & $M_\oplus$ & Observable                & -- \\
Planet radius                    & $r_p$    & --                   & $R_\oplus$ & Observable                & -- \\
Stellar flux                     & $F$      & --                   & W/m$^2$    & Derived                   & -- \\
Planetary albedo                 & $A$      & 0                  & --         & Assumed                   & -- \\
Atmospheric opacity              & $\kappa$ & $5\times10^{-4}$\tablenotemark{a}     & cm$^2$/g   & Assumed                   & \cite{natasha_batalha_resampled_2025} \\
Specific gas constant            & $R$      & 189\tablenotemark{a} & J/kg/K     & Measured\tablenotemark{b} & \cite{chase_janaf_1986} \\
Isobaric heat capacity           & $c_p$    & 849\tablenotemark{a} & J/kg/K     & Measured\tablenotemark{b} & \cite{chase_janaf_1986} \\
Isochoric heat capacity          & $c_v$    & 658\tablenotemark{a} & J/kg/K     & Measured\tablenotemark{b} & \cite{chase_janaf_1986} \\
Surface drag coefficient         & $C_D$    & $3.4\times10^{-3}$   & --         & Derived                   & \cite{wordsworth_atmospheric_2015} \\
Turbulent diffusivity coefficient    & $k_T$    & $8\times10^{-2}$   & --         & Assumed                   & -- \\
Surface integration factor       & $\chi$   & 1                & --         & Derived                   & Appendix \ref{sec:app_xi} \\
Atmospheric integration factor   & $\xi$    & $\pi/4$                & --         & Derived                   & Appendix \ref{sec:app_xi} \\
Surface pressure                 & $p$      & --                   & bar        & Assumed                   & -- \\
\enddata
\tablenotetext{a}{For CO$_2$}
\tablenotetext{b}{Known assuming an atmospheric composition}
\end{deluxetable*}

Planetary mass ($m_p$) and radius ($r_p$) can be determined from radial velocity and transit observations, respectively, and stellar flux ($F$) from stellar observations and modeling. Planetary albedo ($A$) can be estimated from eclipse observations, but can be degenerate with parameters representing the circulation efficiency \citep[e.g.][]{xue_jwst_2024}. Thus, we fix $A=0$ to reduce degeneracies in our fiducial analysis.

The gray gas opacity ($\kappa$) varies depending on the atmospheric composition, as do the specific gas constant ($R$) and the heat capacities ($c_p$, $c_v$). For much of this work, we assume a CO$_2$ atmosphere, as this is expected to be a common abiotic composition from planetary outgassing \citep{elkins-tanton_ranges_2008,cherubim_oxidation_2025}. We take heat capacities from the JANAF Thermochemical Tables \citep{chase_janaf_1986}, and opacities from \texttt{picaso} \citep{natasha_batalha_resampled_2025}. We estimate a gray gas opacity from the Planck mean opacity \begin{equation}
    \kappa\equiv\frac{\int_0^\infty\kappa_\nu(T) B_\nu(T)\dd{\nu}}{\int_0^{\infty}B_\nu(T)\dd{\nu}},
\end{equation}
where $T\approx T_{eq}$. In principle, $\kappa$ could be adjusted based on $\kappa_\nu(T)$ and $B_\nu(T)$ with the solution for atmospheric temperature to achieve a self-consistent result. However, the resolution of the opacity data is poor ($\sim100$ K) at higher temperatures such that this is infeasible, and the effect from $B_\nu(T)$ is negligible enough that we hold the opacity constant. In reality, we expect real gas radiation effects to result in nontrivial deviations from the gray gas behavior, essential for consideration in combination with dynamical effects for a full understanding of planetary energy balance. This assumption is key in enabling a simple analytic picture, however, and we employ it to allow for more focus on the atmospheric dynamics.

We calculate the surface drag coefficient following W15, where 
\begin{equation}
    C_D=\pn{\frac{0.4}{\ln(z/z_0)}}^2,
\end{equation}
where $z_0$ is the roughness height and $z$ is the height of the boundary layer. We choose a typical rocky surface roughness height of 1 cm, and a boundary layer height of 10 m, but note the weak logarithmic dependence on these parameters. The turbulent drag factor is estimated to be about 20 times this value, based on comparison to previous GCM work.

The surface and atmospheric integration factors are order unity factors that do not qualitatively affect our conclusions. Estimates are derived in Appendix \ref{sec:app_xi}. The surface pressure is a key input, as this is one of the variables we aim to constrain from observations. Thus, it is left as a free parameter. Importantly, surface pressures and compositions should only be considered using this model if they result in an approximately optically thin atmosphere.

\subsection{Differences from WTG Methodology}
It is useful to highlight the points in this derivation where the commonly used WTG assumption could have simplified the process. W15 uses this assumption as the starting point of the analogous derivation in Section \ref{sec:thermo}, which allows that model to directly solve for the characteristic wind speed of the system. In \citet{koll_temperature_2016} and \citet{koll_scaling_2022}, wind speeds are estimated using a heat engine scaling, which is intricately related to the WTG assumption in the radiative-convective-subsiding model and the scaling for heat redistribution in \citet{koll_scaling_2022}. The heat engine approach is a useful one, and could merit further exploration, although \citet{koll_temperature_2016} say it is less useful in hot and rapidly-rotating atmospheres, which are of particular interest in this work. Thus, we made the more complicated but illustrative choice to let the wind speed vary. To capture the inherently faster transport of hotter or lighter atmospheres, we scale it with the sound speed $c_s$, calling the factor of proportionality $\alpha$. Our model has only one more input parameter than the \citet{koll_scaling_2022} model, and two more than W15. We explore the parameter space of varying $\alpha$ in Section \ref{sec:alpha_vary}, and the possibilities for scaling $\alpha$ in different regimes in Section \ref{sec:alpha_scaling}.

\section{Results} \label{sec:results}
\subsection{Parameter space exploration}\label{sec:alpha_vary}
We now investigate the behavior of the analytic model developed in Section \ref{sec:model}. First, we performed a parameter space exploration on the effects of changing $\alpha,$ the longitudinal heat transport efficiency. We used as examples three well-studied planets spanning a range of equilibrium temperatures: a tidally locked Earth-instellation planet ($\sim$ 250 K), TRAPPIST-1 b \citep[][$\sim$ 400 K]{agol_refining_2021}, and LHS 3844 b \citep[][$\sim$ 800 K]{vanderspek_tess_2019}. Both exoplanets in this sample were calculated in Section \ref{sec:WTG} to have $\Lambda \leq1$, making them strong testbeds for modeling non-WTG behavior, while the Earth-like planet around an M star would have a rotation period of tens of days, making it a useful reference point for comparing to WTG models.

\begin{figure*}
    \centering
    \includegraphics[width=0.9\linewidth]{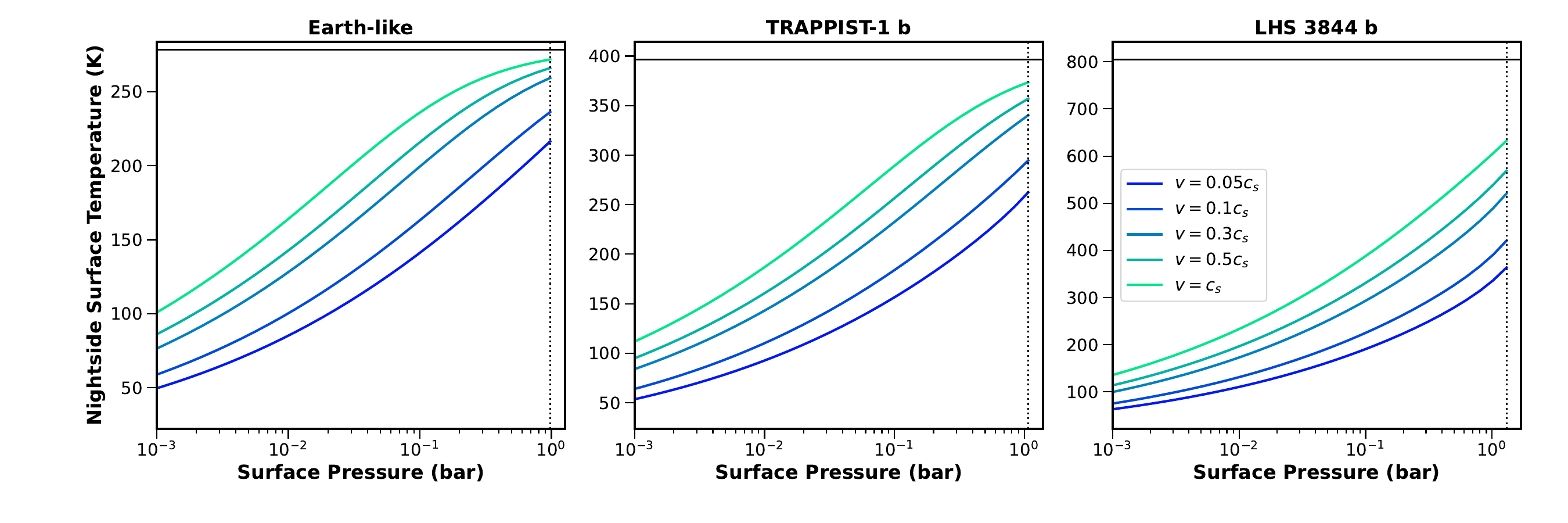}
    \caption{Nightside surface temperature calculations as a function of surface pressure for a gray CO$_2$ atmosphere on four different planets spanning a range of equilibrium temperatures, under the assumptions in \ref{sec:parameters}. Each colored line represents a different value of $\alpha$, the longitudinal wind efficiency. The vertical dotted line represents the $\tau=1$ limit for the optically thin assumption, and the horizontal solid line represents the planet's equilibrium temperature assuming perfect heat redistribution.}
    \label{fig:nightside_surf}
\end{figure*}

\begin{figure}
    \centering
    \includegraphics[width=\linewidth]{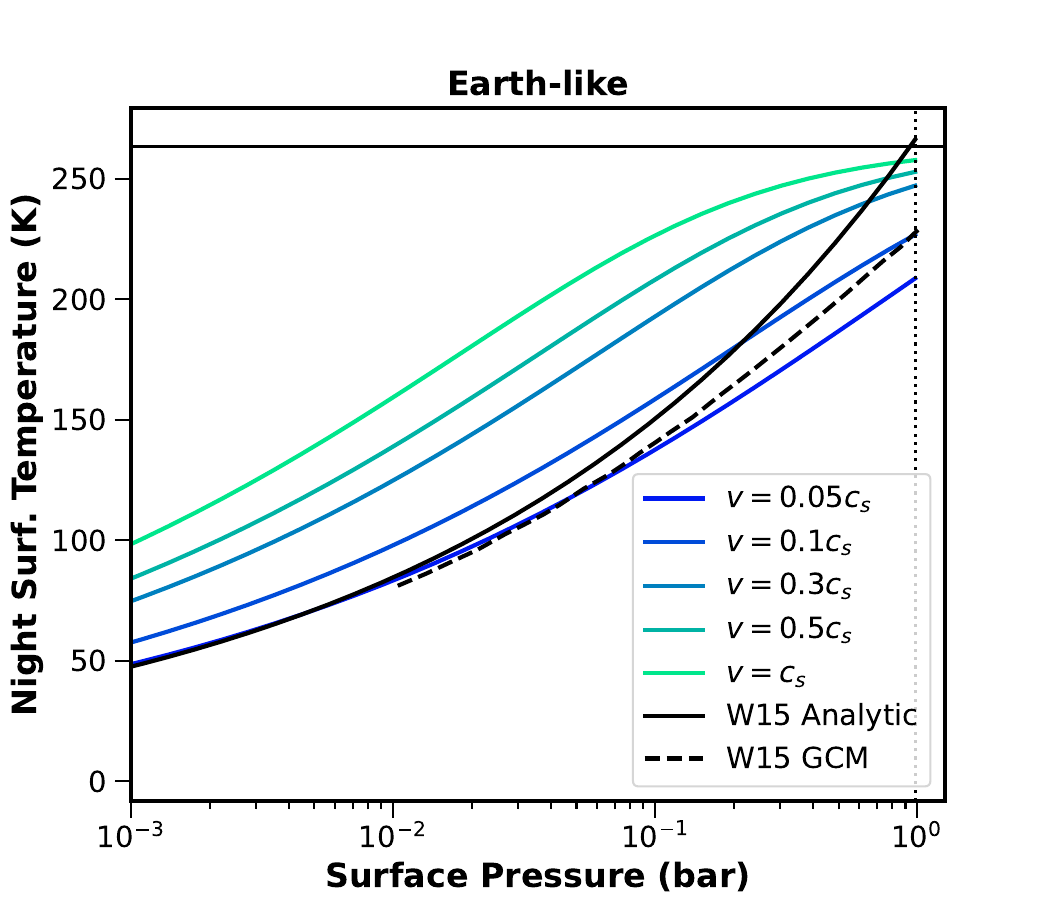}
    \caption{The first panel of Figure \ref{fig:nightside_surf}, with the results of \cite{wordsworth_atmospheric_2015} overplotted in black. The dashed curve represents the analytic model assuming an isothermal WTG atmosphere, and the solid curve represents the results of GCMs run in that work.}
    \label{fig:robin_compare}
\end{figure}

\begin{figure*}
    \centering
    \includegraphics[width=0.9\linewidth]{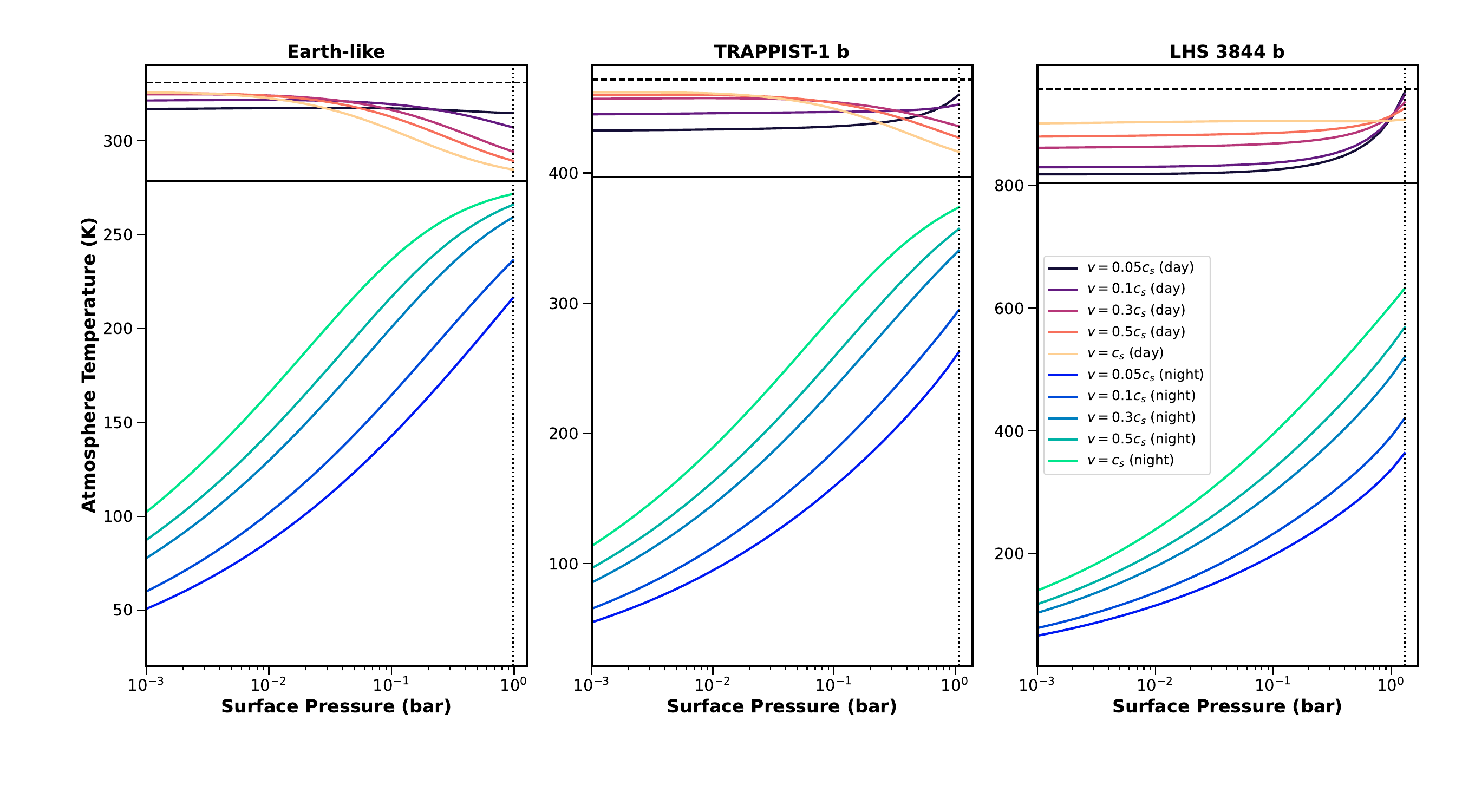}
    \caption{Nightside atmosphere temperatures in blue-greens from the calculations for Figure \ref{fig:nightside_surf}, with dayside atmosphere temperatures shown for comparison in orange-reds. The horizontal dashed line represents the zero-albedo, airless blackbody case ($T_0$ in this work), while the horizontal solid line represents the planet's equilibrium temperature.}
    \label{fig:nightside_atm}
\end{figure*}

\begin{figure*}
    \centering
    \includegraphics[width=0.9\linewidth]{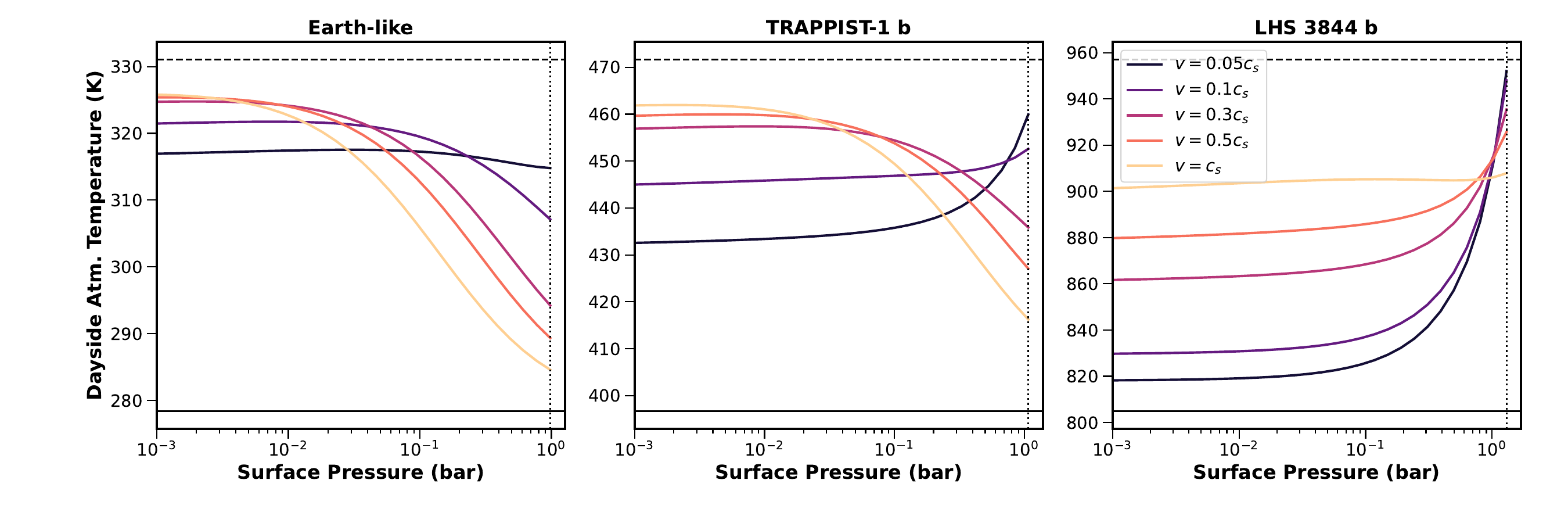}
    \caption{Dayside atmosphere temperatures from the calculations for Figure \ref{fig:nightside_surf}. The horizontal dashed line here represents the zero-albedo, airless blackbody case ($T_0$ in this work), while the solid line still represents the planet's equilibrium temperature.}
    \label{fig:dayside_atm}
\end{figure*}

\begin{figure*}
    \centering
    \includegraphics[width=0.9\linewidth]{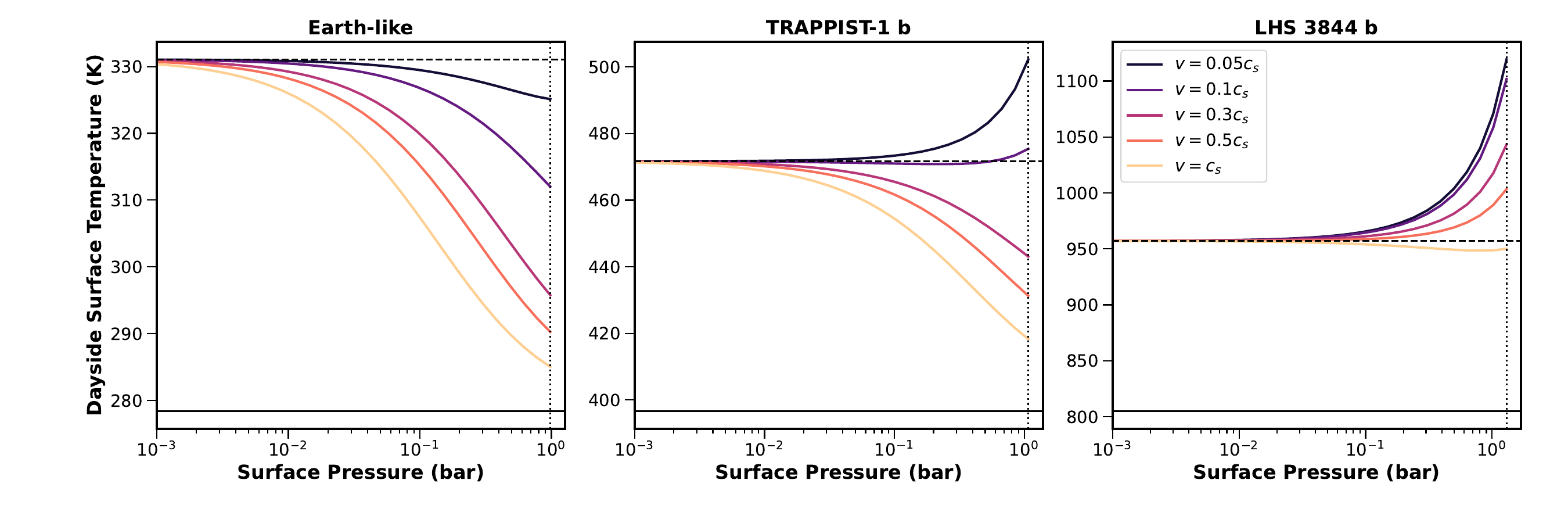}
    \caption{Dayside surface temperatures from the calculations for Figure \ref{fig:nightside_surf}. The horizontal dashed line here represents the surface temperature of a zero-albedo, airless blackbody ($T_0$ in this work), while the solid line still represents the planet's equilibrium temperature.}
    \label{fig:dayside_surf}
\end{figure*}

For each planet, we calculate the dimensionless parameters $\tilde{L},$ $\tilde{C},$ $\tau,$ and $\tilde{V}$ as described in Section \ref{sec:parameters} over a grid of surface pressures and values of $\alpha$. Here, we assume a gray CO$_2$ atmosphere with surface pressure grid spanning from $10^{-3}$ bar up to the $\tau=1$ limit where the optically thin assumption no longer applies (approximately 1 bar for these planets). We test $\alpha$ values over one and a half orders of magnitude, ranging from 0.05 to 1. That is, we let the wind speed vary between the atmosphere's sound speed and one-twentieth of that value. This range is motivated by the GCM comparison in Figure \ref{fig:robin_compare} on the low end, and by the supersonic flows modeled on lava planets in \citet{nguyen_modelling_2020} on the high end. The solutions for each planet are shown in Figures \ref{fig:nightside_surf}, \ref{fig:nightside_atm}, \ref{fig:dayside_atm}, and \ref{fig:dayside_surf}. 

In Figure \ref{fig:robin_compare}, we compare our results with the WTG-regime analytic results of W15, which assumes an isothermal WTG atmosphere on an Earth-insolation tidally locked planet. We find that our model, with $\alpha=0.05$, provides a closer fit and more comparable functional form to the GCMs run in W15, which are run on a slowly-rotating planet and should be well-described by WTG physics. With a sound speed of approximately $240$ m s$^{-1}$, this corresponds to characteristic wind speeds of about 12 m s$^{-1}$, which is quite close to both the GCM wind speeds of about 10-15 m s$^{-1}$ (their Figure 5) and the analytic estimation of about 7 m s$^{-1}$. Thus, this flexible model appears to be consistent with previous WTG results.

The nightside calculations (Figures \ref{fig:nightside_surf} and \ref{fig:nightside_atm}) give identical temperatures within 5 Kelvin between surface and atmosphere, suggesting a strong thermal coupling on the nightside; thus, we also display the dayside atmosphere temperatures in Figure \ref{fig:nightside_atm} for ease of comparison. These are more similar temperatures than predicted from GCMs; W15 finds a tens-of-K difference between the surface temperature and atmospheric temperature on the nightside. This is likely due to the assumption that wind speeds hold over the entirety of the planet, thus causing them to also be high on the nightside, increasing the sensible heat flux to equilibrate the temperatures. It could also be due to the gray gas assumption, as \citet{leconte_3d_2013} found that gray atmosphere models overestimate the nightside surface temperature in comparison to those that incorporate real gas effects. In particular, they find that the nightside absorption of a $\sim$ 1 bar CO$_2$ atmosphere is entirely dominated by the 15$\mu$m band, making the atmosphere particularly non-gray in this regime. In contrast, GCM results show a nightside radiative layer. These differences have a small effect on the day-night contrast, so we do not expect this to affect those results. However, nightside temperature is very important for determining the point of atmospheric collapse, so the nightside dynamics merit further study in this framework.


On all of the planets, as the surface pressure increases, the nightside temperature also increases, as more heat is able to be transferred from the dayside through the atmospheric flow. However, the predicted nightside temperatures vary as a function of the assumed heat transport efficiency, up to 100s of Kelvin over the explored range of transport regimes. For the cooler, Earth-like and TRAPPIST-1 b cases, the nightside temperature comes within a few 10s of Kelvin of the equilibrium temperature for the fastest transport speeds, while the CO$_2$ atmosphere still remains well-described by the optically thin assumption. However, for the hotter LHS 3844 b, an optically thin atmosphere never causes sufficient heat redistribution to reach the predicted equilibrium temperature on the nightside. This is because the amount of energy needed to heat the nightside to the hotter equilibrium temperature of this planet scales faster than the characteristic heat transport speed, even for the most efficient case ($E_\mathrm{therm}\sim c_pT$ versus $c_s\sim\sqrt{T}$). Thus, the result shown by models such as \citep{koll_temperature_2016} and \citep{komacek_atmospheric_2016} that planets with higher equilibrium temperature will on average have higher day-night temperature contrasts is a consequence of this model as well.

On the dayside, at low surface pressures the dayside surface temperatures tend towards the zero-albedo airless blackbody surface temperature, $T_0$, as expected. As the surface pressure increases, however, we observe a variety of behaviors from the model depending on the planet's equilibrium temperature. For the coolest, Earth-like planet in Figures \ref{fig:dayside_atm} and \ref{fig:dayside_surf}, increasing surface pressure of the atmosphere has the effect of cooling both the dayside atmosphere and surface, as predicted in models such as \cite{koll_scaling_2022}, hereafter K22. The cooling is more efficient for higher values of $\alpha$.

For planets with hotter equilibrium temperatures such as TRAPPIST-1 b and LHS 3844 b, increasing surface pressure of an optically thin atmosphere no longer necessarily decreases the dayside surface temperature, as assumed in models like K22. If heat transport is efficient, this will happen for intermediate surface pressures ($\sim0.1$ bar), but as the atmosphere approaches the optically thick limit, the decreased radiative timescale of the hotter planet will lead to an increased temperature, even potentially raising the surface temperature above the zero-albedo airless limit. Although these model results are just for the surface temperatures, not taking into account the contribution from the atmosphere, observations of some hot rocky planets are showing eclipse depths suggestive of dayside temperatures hotter than predicted for a bare rock \citep{valdes_hot_2025,fortune_hot_2025,xue_jwst_2025}. For less efficient heat transport ($\alpha \lesssim 0.1$ for TRAPPIST-1 b and $\alpha \lesssim 0.5$ for LHS 3844 b), $T_{s,d}$ can even monotonically increase with atmospheric surface pressure. The dayside atmosphere also experiences this effect, leading to a transition from a sensible heat flux dominated regime where more efficient transport causes higher $T_{a,d}$, to a radiation-dominated regime where less efficient transport allows for more heat transfer from the surface.

\begin{figure*}
    \centering
    \includegraphics[width=0.7\linewidth]{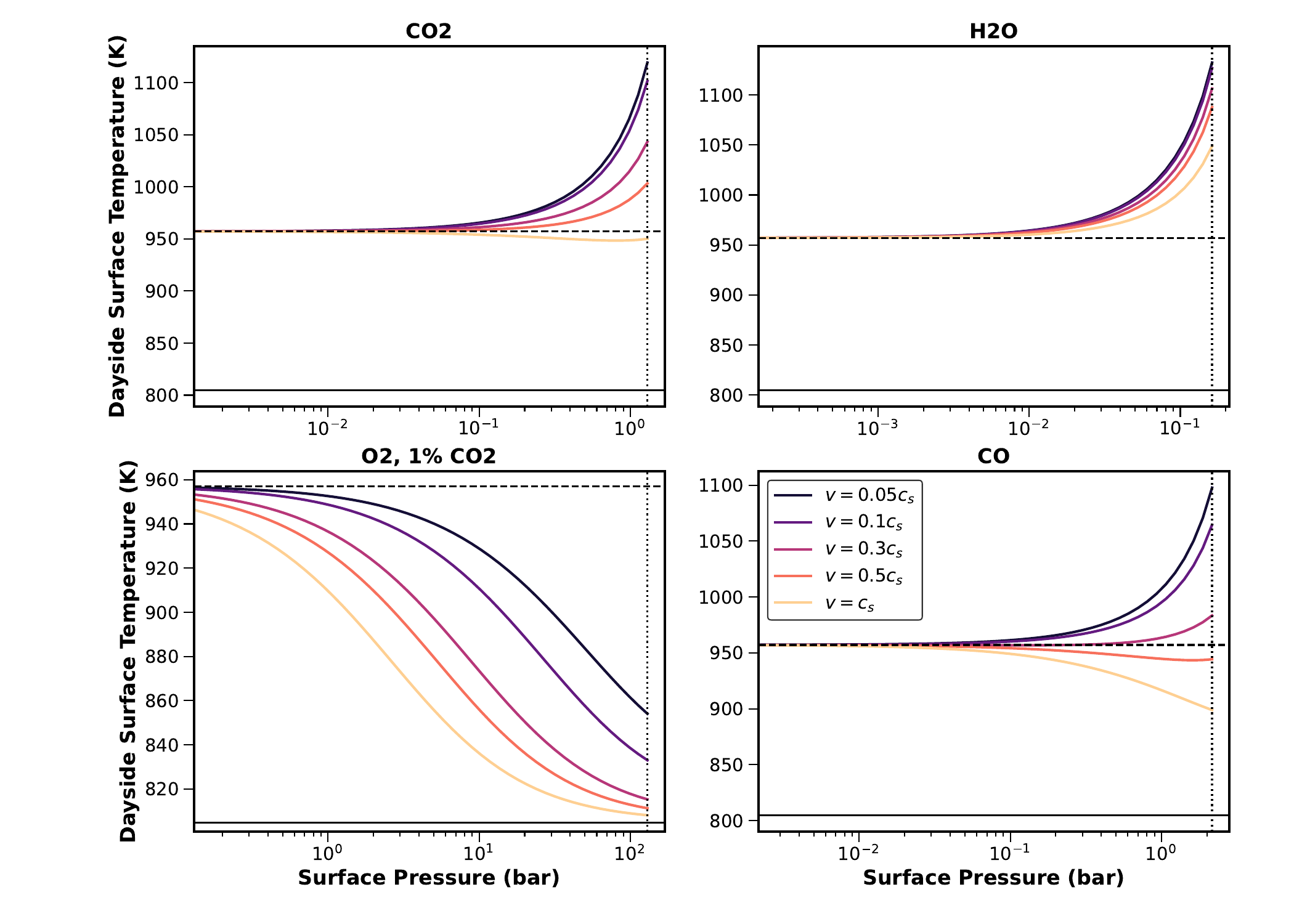}
    \caption{Dayside surface temperatures for LHS 3844 b under different gray optically thin atmospheric compositions. Each colored line represents a different value of $\alpha$, the longitudinal wind speed. The horizontal dashed line represents the surface temperature of a zero-albedo, airless blackbody ($T_0$ in this work), and the vertical dotted line represents the $\tau=1$ limit for each composition. Note the different ranges on the pressure axis in each panel, representing different pressure regimes for which this calculation is most applicable.}
    \label{fig:comp_LHS}
\end{figure*}

We also explore the effects of atmospheric composition in this model. Figure \ref{fig:comp_LHS} shows the result of model evaluations for LHS 3844 b under four different atmospheric compositions: CO$_2$, H$_2$O, O$_2$ with 1\% CO$_2$, and CO. When compared to CO$_2$, H$_2$O absorbs more strongly in the mid-IR thermal emission band of the planets considered. Thus, we find the greenhouse-like heating of the dayside surface to occur for lower surface pressures for the steam atmosphere than for the carbon dioxide atmosphere, and to occur for all modeled transport efficiencies. Additionally, the atmosphere reaches an approximate optical depth of 1 at an order of magnitude lower surface pressures (about 0.2 bar) for water than for CO$_2$, limiting the application of this model to those atmospheres. CO, with its lighter molecular weight and lower mid-IR gray opacity, is able to redistribute more heat than CO$_2$. It behaves qualitatively similarly, with less efficient heat redistribution resulting dayside greenhouse heating as surface pressure increases and more efficient heat redistribution causing dayside cooling as surface pressure increases, but the dayside is ~50 Kelvin cooler with a CO atmosphere than with a CO$_2$ atmosphere across all cases.

On the other hand, an atmosphere composed mostly of weakly absorbing O$_2$ with a small fraction of CO$_2$ serves to exclusively cool the dayside. While the radiative effect of this atmosphere in the surface energy balance is small, the effect of the sensible heat flux and day-night transport is large compared to other atmospheric compositions, especially at the large surface pressures for which it remains approximately optically thin. At surface pressures approaching $10^2$ bar, it is likely that the vertical structure of the atmosphere will play an increasing role in the dynamics, potentially limiting the applicability of this model. Further consideration of massive optically thin atmospheres could be an interesting area of future study. Additionally, real gas radiation effects differ for different compositions; some gases (such as H$_2$O) are better-approximated by a gray model than others (such as CO) \citep{pierrehumbert_principles_2010}.

\subsection{Scaling of $\alpha$ and dynamical regimes}
\label{sec:alpha_scaling}
The key free parameter in these calculations and results is $\alpha=\frac{\abs{\mathbf{v}}}{c_s}$, which has a large qualitative effect on the behavior of the system. Thus, constraining this value is key to making predictions about exoplanet atmospheres. In a real planetary atmosphere, the longitudinal wind speed is set by a variety of factors and is not necessarily a constant fraction of the sound speed. Previous work has estimated wind speeds in a variety of ways, including the usage of the comparison of planetary atmospheres to heat engines \citep{koll_temperature_2016,koll_scaling_2022}, or through a numerical perturbation method \citep{song_influence_2025}. 

In any case, for a tidally locked planet the bulk flow is forced by the pressure gradient in the atmosphere caused by the difference in heating between day and night. This force can be balanced by a variety and a combination of factors, which correspond to different regimes of a planet's dynamics. We estimate the scaling of $\alpha$ for a planet that is well-described by the WTG assumption, and for one that is not in this regime and where instead rotation plays a dominant role in the dynamics (a rapid rotator). We follow \citet{komacek_atmospheric_2016} and \citet{zhang_effects_2017} in this calculation. Other regimes exist, such as planets dominated by surface drag or Ohmic drag \citep{rauscher_three-dimensional_2013,vigano_inflated_2025}, but previous terrestrial results have found that planetary rotation is the first-order effect that causes deviations from WTG physics \citep{koll_temperature_2016}, so we neglect their consideration in this analysis. 

The momentum equation for a planet's atmosphere, neglecting drag terms, can be written as 
\begin{equation}
    \pdv{\mathbf{v}}{t}+\mathbf{v}\cdot\mathbf{\nabla}\mathbf{v}=-\frac{\mathbf{\nabla}p}{\rho}+f\hat{k}\cross\mathbf{v}+\nu\nabla^2\mathbf{v}.
\end{equation}
Here $\mathbf{v}$ represents the wind velocity, $p$ is the atmospheric pressure, $\rho$ is the atmospheric density, $f$ is the planet's rotational frequency, $\hat{k}$ is the rotational axis, and $\nu$ is the atmospheric viscosity. For a steady-state atmosphere, we have $\pdv{\mathbf{v}}{t}\sim0.$ 

If we assume that viscosity is negligible, then the pressure gradient forcing from the difference in day-night heating must be balanced by horizontal advection and the Coriolis force. For a slowly rotating planet, the Coriolis term is negligible, and we have $-\frac{\nabla p}{\rho}\approx\mathbf{v}\cdot\nabla\mathbf{v}$. This is the WTG regime, in which the characteristic wind speed scales as \begin{equation}
    \frac{R\Delta T}{r_p}\sim\frac{U^2}{r_p},
\end{equation}
taking the characteristic length scale to be the planetary radius, which we can simplify to solve for $\alpha$:
\begin{equation}
    \alpha c_s\sim\sqrt{R(T_{a,d}-T_{a,n})}
\end{equation}
\begin{equation}
    \alpha^2 R T_{a,d}\sim\sqrt{R(T_{a,d}-T_{a,n})}.
\end{equation}
\begin{equation}
    \alpha\sim\sqrt{1-\tilde{T}_2}.
\end{equation}
This scaling captures the principle of the WTG regime where increasing day-night temperature contrasts are balanced by faster heat transport---as $\tilde{T}_2=\frac{T_{a,n}}{T_{a,d}}$ decreases, $\alpha$ increases. If there were to be a very hot dayside and very cold nightside $(\tilde{T}_2\to0)$, then the atmospheric velocities would approach the sound speed. However, since $\alpha$ governs the amount of heat brought to the nightside, increasing it increases $\tilde{T}_2$, allowing the system to converge to a subsonic flow.

If the planet is rotating more rapidly, then the Coriolis effect is no longer negligible. For a planet where this force dominates, we will have $-\frac{\nabla p}{\rho}\approx f\hat{k}\cross\mathbf{v}$. This scales as
\begin{equation}
    \frac{R\Delta T}{L}\sim2\Omega U
\end{equation}
which we can again solve for $\alpha$:
\begin{equation}
    \alpha\sim\frac{R\Delta T}{2\Omega L c_s}
\end{equation}
\begin{equation}
    \alpha^2\sim\frac{R^2(T_{a,d}-T_{a,n})^2}{4\Omega^2L^2\gamma R T_{a,d}}\sim\frac{RT_0\tilde{T}_0\tilde{T}_1(1-\tilde{T}_2)^2}{4\Omega^2L^2\gamma},
\end{equation}
where $\gamma$ is the adiabatic index.
We can relate this to the parameter $\Lambda$ from Section \ref{sec:WTG}, giving us
\begin{equation}
    \alpha\sim 2^{1/4}\gamma^{-1}(\tilde{T}_0\tilde{T}_1)^{1/2}\Lambda(1-\tilde{T}_2).
\end{equation}
We see that $\alpha$ similarly increases with increasing day-night contrast, but the extra factor of $\Lambda$, which should be less than 1 for planets in this regime, will reduce the maximum possible transport speed. This represents a simple model for the reduction of the heat transport efficiency on planets where the WTG assumption is less applicable.

With these scalings, we can solve the equations from Section \ref{sec:model} without the assumption of a value of $\alpha$.
\begin{figure*}
    \centering
    \includegraphics[width=0.9\linewidth]{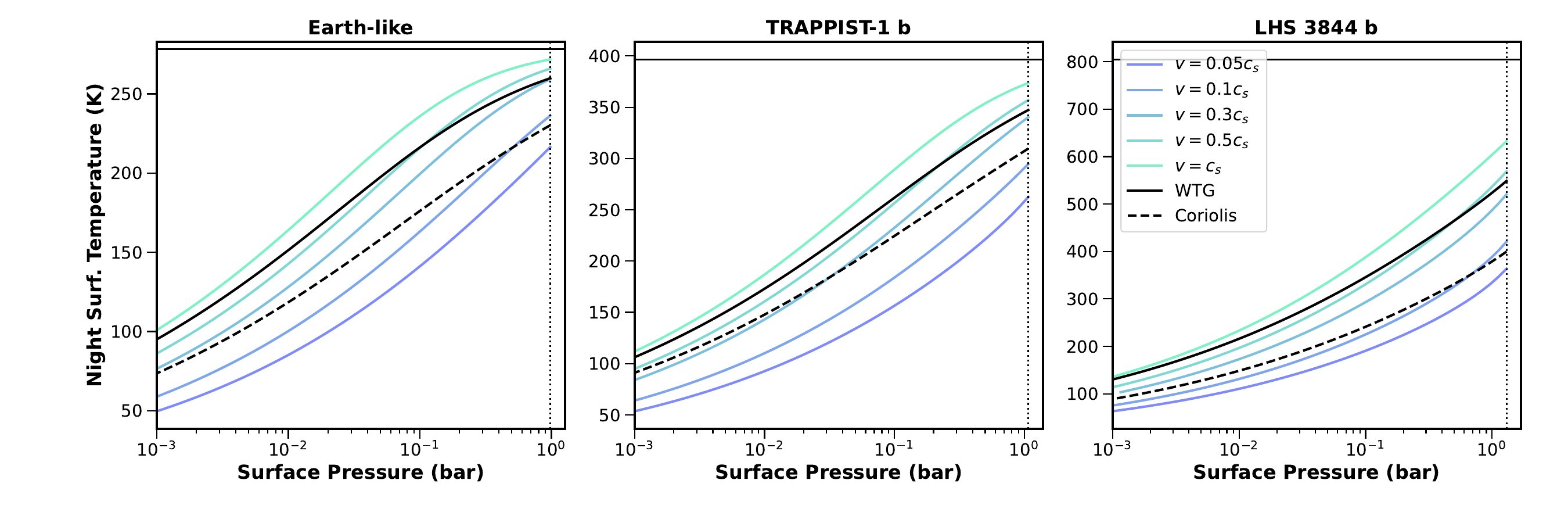}
    \caption{Nightside surface temperatures predicted by self-consistent models of four planets, under the assumption of the WTG regime (solid black line) and a Coriolis-dominated circulation (dashed black line). The results of Figure \ref{fig:nightside_surf} where constant values of $\alpha$ are assumed are shown in the transparent blue-green lines. The solid horizontal lines represents the planet's equilibrium temperature.}
    \label{fig:consistent_night}
\end{figure*}
\begin{figure*}
    \centering
    \includegraphics[width=0.9\linewidth]{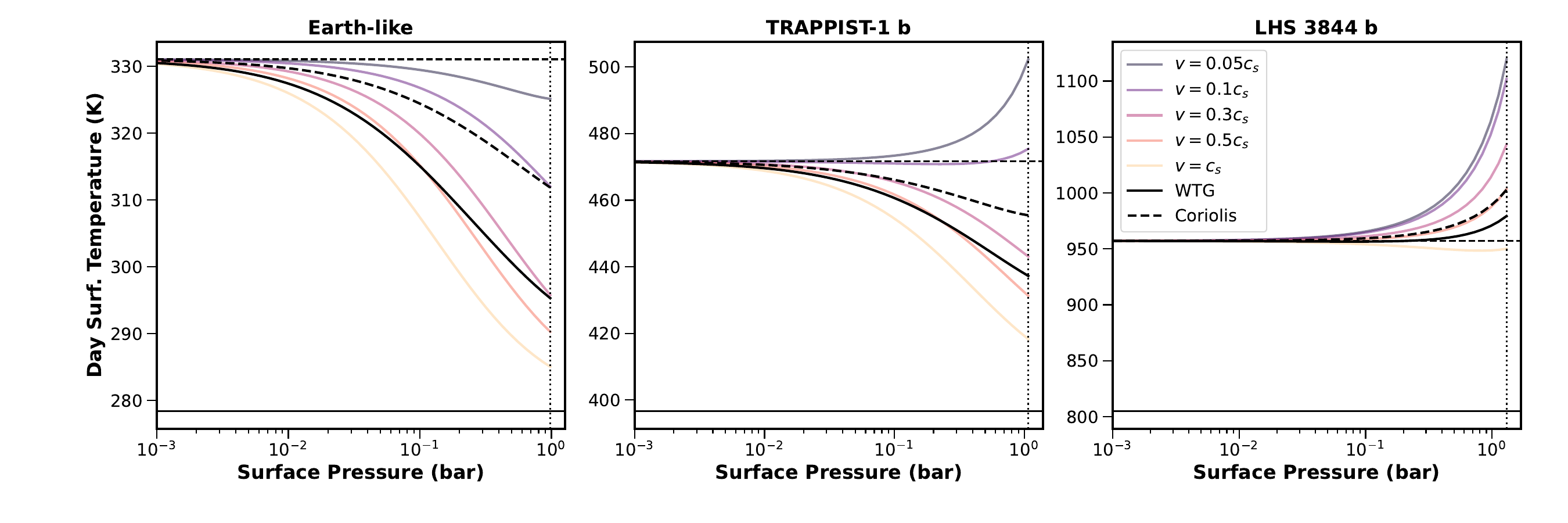}
    \caption{Dayside surface temperatures predicted by self-consistent models of four planets, under the assumption of the WTG regime (solid black line) and a Coriolis-dominated circulation (dashed black line). The results of Figure \ref{fig:dayside_surf} where values of $\alpha$ are assumed are shown in the transparent red-purple lines. The solid horizontal line represents the planet's equilibrium temperature, and the dashed horizontal line represents the surface temperature of a zero-albedo, airless blackbody ($T_0$).}
    \label{fig:consistent_day}
\end{figure*}
Figures \ref{fig:consistent_night} and \ref{fig:consistent_day} show the results of these calculations compared with calculations from above for which $\alpha$ was held constant. For these calculations, the rotation period of the Earth-like planet was set to 1 day, because the slow rotation period of a tidally locked Earth-like planet would make the Coriolis-dominated case meaningless. The Coriolis-dominated regime represents less day-night heat transport than the WTG regime for all planets, and the Coriolis-dominated and WTG regimes are closest for TRAPPIST-1b, which has the highest $\Lambda$ value of the three planets considered. Additionally, as $p_\mathrm{surf}$ increases, heat transport does not need to be as efficient to balance the pressure gradient term and so the self-consistent solution moves from constant-$\alpha$ lines with higher $\alpha$ to those with lower $\alpha$. Similarly, because of the larger day-night temperature differences of hotter planets, the self-consistent solutions for hotter planets line up with higher $\alpha$ values than for cooler planets. 


\section{Discussion} \label{sec:discussion}
We developed a model to provide first-principles order-of-magnitude estimates of rocky planet temperatures as a function of atmospheric composition and thickness over different planetary rotational regimes. This model builds on the work of \citet{pierrehumbert_palette_2011},\citet{yang_low-order_2014}, and \citet{wordsworth_atmospheric_2015}, but relaxes the Weak Temperature Gradient assumption used in these and other models to provide a more flexible framework to accommodate the hot, rapidly rotating terrestrial planets currently being observed with the James Webb Space Telescope.

\subsection{Observational constraints}
We can now determine what we expect for observations of planets both in and out of the WTG regime by translating the model results to observables. Namely, we want to find the expected planetary brightness temperatures of the day and night side under different model assumptions. These can be estimated from secondary eclipses and phase curves, and represent key pieces of evidence in our understanding of exoplanet atmospheric dynamics.

In doing this conversion, we need to take into account the viewing geometry of the observer, as well as the combined effect of both atmosphere and surface. We calculate the brightness temperature for the day and night under the optically thin atmosphere assumption using
\begin{equation}
    T_b^4=T_s^4+2\tau(T_a^4-T_s^4).
\end{equation}
This takes into account the integrated airmass of the thin atmosphere via the factor of 2. The viewing angle effect is derived in Appendix \ref{sec:app_obs}, where we find that the brightness temperature on the dayside should be multiplied by a factor of $(4-\tilde{T}^4_2)/3$, where $\tilde{T}_2$ is defined in Section \ref{sec:model} to be the mean nightside atmospheric temperature over the mean dayside atmospheric temperature. This factor is derived thus mainly for comparison of these results with those of K22, which assumes a similar weighting effect; future work could explore a more physically motivated calculation of this effect, particularly in light of the hot dayside surface temperatures predicted in this analysis.

\begin{figure*}
    \centering
    \includegraphics[width=0.9\linewidth]{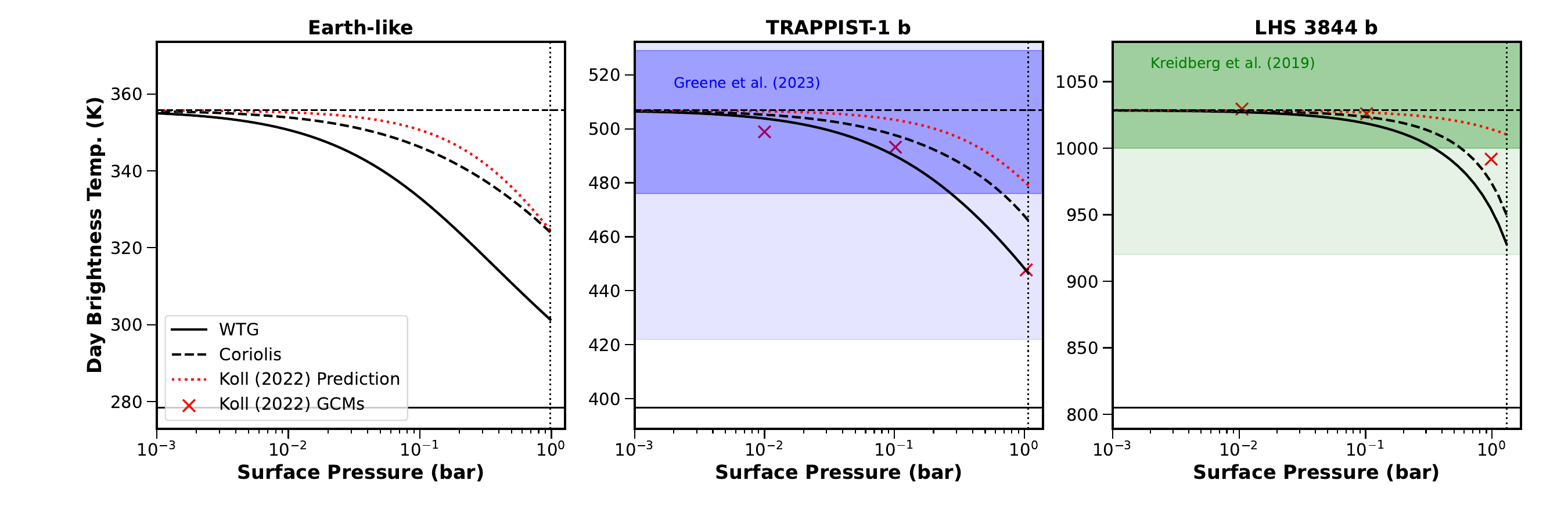}
    \caption{Dayside brightness temperature for the self-consistent models solved in Figures \ref{fig:consistent_night} and \ref{fig:consistent_day}. The solid line represents a WTG-regime circulation and the dashed line represents a Coriolis-dominated circulation. Horizontal and vertical lines are the same as in Figures \ref{fig:dayside_atm} and \ref{fig:dayside_surf}. The red crosses represent the results of general circulation models run in K22, for the $\tau\propto p_s$ case, and the red dotted curve represents the scaling relationship derived in that work. The dark shaded regions represent the 1-$\sigma$ confidence intervals on the dayside temperatures of the planets, by \citet{greene_thermal_2023} for TRAPPIST-1 b and by \citet{kreidberg_absence_2019} for LHS 3844 b, and the light shaded regions represent the 3-$\sigma$ confidence intervals.}
    \label{fig:brightness_day}
\end{figure*}

\begin{figure*}
    \centering
    \includegraphics[width=0.9\linewidth]{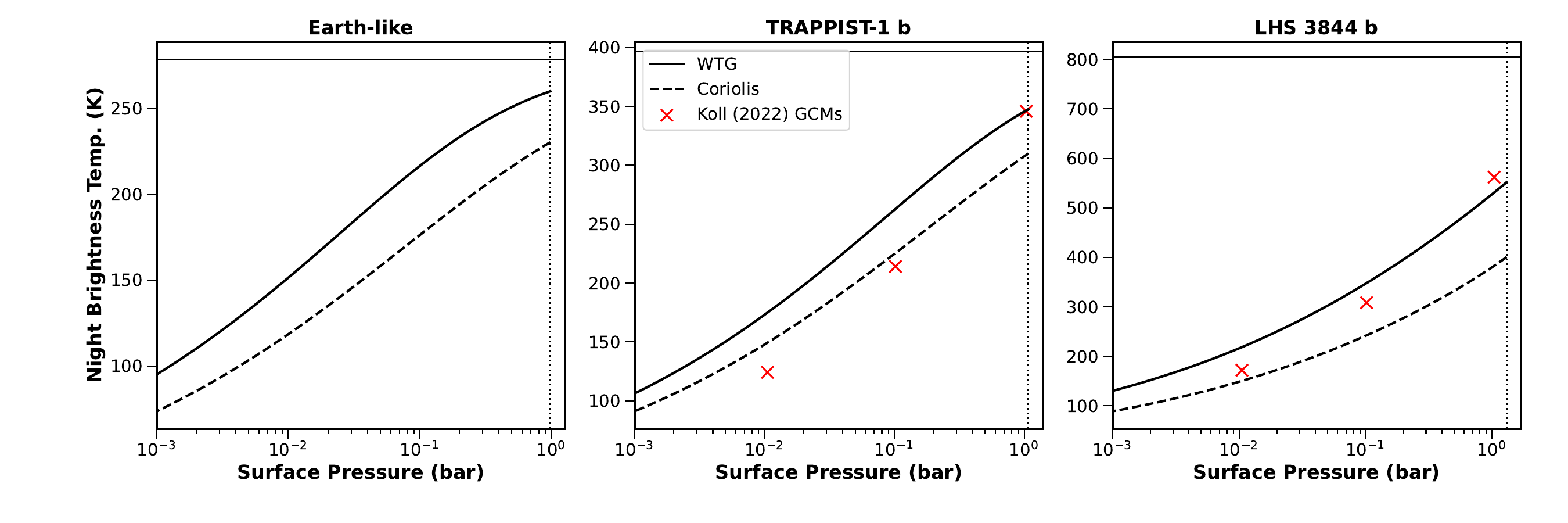}
    \caption{Estimated nightside brightness temperature for the self-consistent models solved in Figures \ref{fig:consistent_night} and \ref{fig:consistent_day}. The solid line represents a WTG-regime circulation and the dashed line represents a Coriolis-dominated circulation. Horizontal and vertical lines are the same as in Figures \ref{fig:dayside_atm} and \ref{fig:dayside_surf}. The red crosses represent the results of general circulation models run in K22, for the $\tau\propto p_s$ case.}
    \label{fig:brightness_night}
\end{figure*}

With calculated brightness temperatures we can now compare directly to observations. To illustrate the potential use of this model in quantifying possible atmospheres, we show the 1-$\sigma$ and 3-$\sigma$ JWST dayside temperature constraints on TRAPPIST-1 b from \citet{greene_thermal_2023} on Figure \ref{fig:brightness_day}. We see that no optically thin CO$_2$ atmospheres can be ruled out from energy balance considerations alone, as all fall within the 3$\sigma$ constraints for both planets, regardless of rotational regime. Results such as \citet{greene_thermal_2023} and \citet{ih_constraining_2023} rule out CO$_2$ atmospheres with pressures above $10^{-2}$ bar, but take advantage of the high CO$_2$ absorption around the 15 micron band of these observations, which would further reduce the eclipse depth if present, beyond just the effects of heat redistribution. In the same way, they place limits on the presence of trace amounts of CO$_2$ in thick atmospheres, ruling out 1\% CO$_2$ in O$_2$ atmospheres of 0.1 bar and above, while the effect of heat redistribution alone in our model is unable to place constraints on these types of atmospheres, even for at surface pressures higher than 1 bar.

Importantly, these analyses of TRAPPIST-1 b and similar ones for LHS 3844 b \citep[e.g.][]{kreidberg_absence_2019,whittaker_detectability_2022}, although they simulate detailed emission spectra, still rely on analytic heat redistribution models, specifically that of K22. Thus, models of this type and their assumptions can bias constraints on atmospheric composition and surface pressure. As an example, we imagine that the constraints on the dayside temperatures of TRAPPIST-1 b and LHS 3844 b were to improve such that the 1-$\sigma$ shaded regions in Figure \ref{fig:brightness_day} became statistically conclusive. We would find from the energy balance models presented in this work, that under the WTG assumption, values above $0.1$ bar of a CO$_2$ atmosphere would be ruled out, while for a Coriolis-dominated atmosphere we could only rule out values above $0.5$ bar. Similarly, the LHS 3844 b dayside temperature constraint would be inconsistent with $4\times10^{-2}$ bars of CO$_2$ under our WTG model, but only inconsistent with values above $2\times10^{-1}$ bars for the Coriolis-dominated model. This difference in inferences based on the dynamical regime would be propagated to any 1-dimensional models used for direct comparison with eclipse measurements as well, and could be further exacerbated by effects such as the weighting of the emission over the planetary disc under these different regimes.

In principle, a detection or nondetection of nightside emission would allow for similar inferences to be made, although this is more difficult because of lower fluxes. The \citet{kreidberg_absence_2019} nondetection of LHS 3844 b's nightside emission is not even 1-$\sigma$ inconsistent with any thin CO$_2$ atmospheres given this model.

\subsection{Comparison to other models}
The scaling derived by K22 is widely used and has had great success in predicting first-order behavior of a wide range of atmospheric pressures and compositions in a simple framework. However, the Koll model relies on the use of the WTG assumption, and assumes an ansatz functional form for the behavior of the system as the thickness of the atmosphere is increased. We have shown in this work that the WTG regime may not be fully applicable to the population of planets currently being observed by JWST, and thus the model we present here could be an improvement in these cases. It is still possible in our model's framework to use WTG physics, which could be useful as observations push towards cooler, slower-rotating planets in search of habitable environments. The use of a prescribed functional form for the effect of increasing surface pressure in K22 could additionally bias conclusions from the use of their model, particularly in the intermediate-pressure regime where the specifics of the relationship would make the largest difference. In particular, the assumption that the emission from planets with extremely thick atmospheres should necessarily approach the uniform-redistribution limit is inconsistent with extensive study of hot Jupiters. 

The Koll analytic model does largely match the general circulation models run in that paper, which do not use the WTG assumption as they directly model the fluid dynamics of the atmosphere. However, the GCMs from K22 rely on gray gas radiative transfer, which misses pieces of physics key to the global energy balance and circulation \citep{pierrehumbert_principles_2010}, so they are less useful for comparing our model and that of K22 to a realistic atmosphere. Our analytic results qualitatively agree with GCM studies that used more complex radiative transfer models, such as \citet{selsis_thermal_2011}, as do the analytic results of K22, but more work is necessary to fully understand the effects of this simplification.

We compare brightness temperatures calculated using our model to those from the GCM runs in K22 in Figure \ref{fig:brightness_day} and \ref{fig:brightness_night}. Our first-principles simple model is able to reproduce the major features of these results, matching the brightness temperatures within 5\%. Also compared are the analytic predictions from the WTG analytic model presented in that work, which reproduce the GCM results on a similar level. Interestingly, the K22 analytic model predicts hotter daysides for these planets than the model presented in this work, although it also predicts hotter daysides than the GCMs as well. This is likely due to the differences in the assumed weighting of the observed emission towards the hot substellar point for thin atmospheres between these models, as well as the choice to model the energetics of the dayside as a whole in the Koll model as opposed to modeling the dayside surface and atmosphere individually in this model. While this choice adds complexity to our model, especially in the calculation of the total emission from surface and atmosphere, our model only has one more input parameter than K22 and two more than W15, and this choice increases its predictive power. Effects localized to the surface such as melting or effects localized to the atmosphere such as the formation of clouds can be incorporated more easily in future work.

\subsection{Limitations and future work}
Because GCMs are able to use fewer simplifying assumptions, the role of analytic models is to provide physical motivation and interpretability. This new model represents a step in that direction, but there are further steps to take in the future.
For one, this work would benefit from a dedicated companion GCM study, an approach we are pursuing at this time. It would be of particular interest to explore a planet's three-dimensional temperature structure and how it relates to the observed emission, especially over different rotational regimes, like the work of \citet{hammond_rotational_2021}. Vertical temperature structure has an important role in atmospheric circulation \citep{watkins_gravity_2010,read_dynamics_2011,koll_temperature_2016}, so the lack of consideration of this dimension in this work is a limitation that could be explored further. Similarly, the assumption of a uniform surface pressure over the planet is also worth further study, as the overturning circulation may decrease the surface pressure, and thus atmospheric mass, on the dayside (shown for example in the GCMs of W15 at the 5\% level). We briefly pursued including this affect in our model, but did not find enough of a difference from the results presented here to justify the added complexity. However, these effects, when combined with the observation angle weighting of the dayside, may be important to consider in future work.

The gray atmosphere assumption is a major limitation of this model, as spectral windows can non-negligibly affect the radiative transfer in exoplanet atmospheres \citep{cmiel_characterizing_2025}. Beyond this, the assumption that the atmospheric optical depth would scale linearly with surface pressure may break down, especially as surface pressure increases \citep{pierrehumbert_principles_2010}. GCM work with full radiative transfer could be used to explore this further, particularly as a function of location on the planet and input stellar spectrum, but Section 4 of K22 could provide a useful starting point for further analytic study. When incorporating  non-gray gas effects, we might expect a CO$_2$ atmosphere to transport more heat away from the dayside than found in this model, because the 15-micron feature could allow the atmosphere to absorb efficiently from the dayside surface and radiate efficiently to the nightside surface, while spectral windows could allow for direct radiation from the nightside surface to space \citep{wordsworth_atmospheric_2015}. Of course, this behavior depends on temperature, so it would be worth exploring the planets for which these effects would be strongest, as well as the variety of potential atmospheric compositions and their different behaviors.

Similarly, molecular opacities, heat capacities, and other substance properties depend nontrivially on temperature. We should improve our measurements of these properties for a greater variety of species at a greater variety of temperatures, and explore their effects on atmospheric systems.

An additional complication to the constraint on atmospheric condition from exoplanet temperature structures is the presence of aerosols. Clouds or hazes on the dayside or nightside have the potential to both increase and decrease day-night temperature contrasts \citep[e.g.][]{yang_stabilizing_2013,powell_nightside_2024}. The coupled effect of the atmospheric temperature on the condensation of various species and the radiative effects of aerosols on temperature structures leads to a wide variety of possible effects. It would be interesting to attempt an expansion of this framework to model these effects.

Finally, the boundary layer physics is an uncertainty in our model that could use further investigation. Especially on the nightside, the specifics of the vertical temperature structure and boundary layer are of key importance \citep{wordsworth_atmospheric_2015,joshi_earths_2020}. In particular, a nightside temperature inversion near the surface can increase the likelihood of atmospheric collapse on the nightside, limiting the range of possible atmospheres at low pressures. This could have implications for atmospheric retrievals interested in constraining the presence of low-pressure atmospheres. The background atmospheric flow will also very strongly as a function of planetary rotation rate \citep{haqq-misra_demarcating_2018}, making this particularly interesting to understand in our framework.

The specifics of the boundary layer largely affect the model presented in this paper through the surface drag and turbulent diffusivity parameters, which have a non-negligible effect on the results presented here. We estimated the turbulent diffusivity from comparison to previous GCMs, but it should be possible to explore this more self-consistently through eddy diffusion and Ekman layer models or Monin-Obhukov theory.

\section{Conclusions}
\label{sec:conclusions}
In this paper, we present a physically motivated analytic model of the atmospheres of tidally locked terrestrial planets, designed to calculate temperatures of a planet's dayside and nightside atmosphere and surface as a function of potential atmospheric compositions and thicknesses. We find that a key assumption in previous models, the Weak Temperature Gradient assumption, likely does not apply to over 40\% of the planets currently being observed with JWST, owing to the increased importance of the planetary rotation on the atmospheric circulation. 

We develop a new model without the use of this assumption, which assumes a steady-state, gray, optically thin atmosphere. We allow the longitudinal heat transport speed of this atmosphere to be a free parameter, as well as deriving scalings for this value in different dynamical regimes, and find that a planet's nightside temperature can vary on the order of 100s of Kelvin depending on the dynamical regime of the atmosphere. We also observe that the dayside temperature can display degeneracies between surface pressure and temperature, with temperatures at or above that of a bare rock surface under a number of plausible atmospheric compositions and transport efficiencies due to greenhouse warming from the atmosphere.

We compare our results with general circulation models from the literature for planets that should be well-described and poorly-described by WTG physics, and find excellent agreement (within 5\% of predicted surface temperatures). We show how this model can be used to assess the plausibility of various atmospheres from observations, and give a demonstration that assumptions on atmospheric dynamics can bias our constraints on these atmospheres at the order-of-magnitude level.

As models of exoplanet atmospheres continue to grow in complexity and resolution, it is worthwhile to continue to pursue simple, interpretable approaches to evaluate assumptions and enhance our understanding. This model represents a demonstration of the flexibility, speed, and utility that analytic work can bring to our understanding of these other worlds.

\begin{acknowledgments}
We thank Thaddeus Komacek for insightful discussions that helped improve this work, and the anonymous reviewer for their detailed and worthwhile comments. This research has made use of the NASA Exoplanet Archive, which is operated by the California Institute of Technology, under contract with the National Aeronautics and Space Administration under the Exoplanet Exploration Program.

\end{acknowledgments}

%





\appendix
\section{Estimation of integration constants}
\label{sec:app_xi}
\begin{figure}
    \centering
    \includegraphics[width=0.5\linewidth]{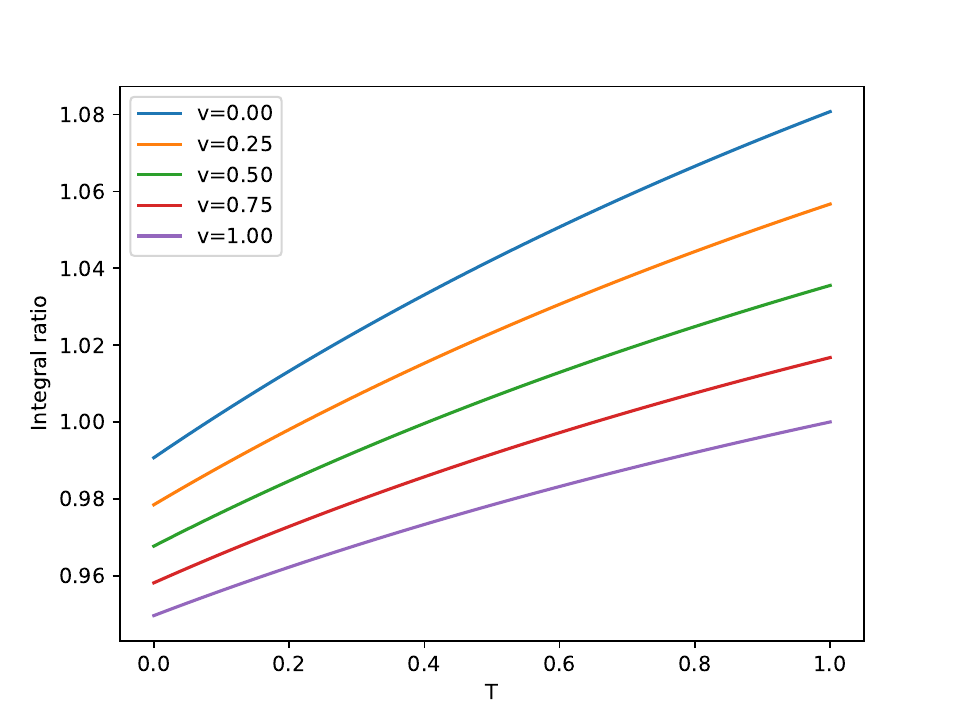}
    \caption{Ratio of the integrals for $\xi$ in Appendix \ref{sec:app_xi}, which is approximately 1 over a large region of parameter space.}
    \label{fig:app_xi}
\end{figure}

We first turn our attention to $\xi$, defined in Section \ref{sec:model} such that
\begin{equation}
    \overline{\abs{\mathbf{v}}T_{a,d}(\Delta T_a)}=\xi\overline{\abs{\mathbf{v}}}\overline{T_{a,d}}
    \left(\overline{T_{a,d}}-\overline{T_{a,n}}\right);
    \label{eq:xi_1}
\end{equation}
that is, $\xi$ captures the effects of weighting and spherical-geometric effects in the dayside average of the advective term of Equation \eqref{eq:localatm}. To estimate this value, we assume an example functional form for $T_a$, which has maximum $T_{a,\max}$ at the substellar point and $T_{a,\min}$ at the antistellar point. Between these values, it is proportional to the cosine of the substellar latitude $\theta_z$ (here $\theta$ for simplicity) such that
\begin{equation}
    T_a=\frac{T_{a,\max}-T_{a,\min}}{2}\cos\theta+\frac{T_{a\max}+T_{a,\min}}{2}.
\end{equation}
With this, we can take
\begin{equation}
    \Delta T_a\approx\frac{T_{a,\max}-T_{a,\min}}{2}\sin\theta.
\end{equation}
We also assume that $\mathbf{v}$ points in the direction of increasing $\theta_z$, carrying heat from day to night, and that it peaks near the terminator, so we take
\begin{equation}
    \abs{\mathbf{v}}=(v_{\max}-v_{\min})\sin\theta+v_{\min}.
\end{equation}
Now, the integral on the left of equation \eqref{eq:xi_1} becomes
\begin{multline}
    \frac{1}{2\pi r_p^2}\iint_d \abs{\mathbf{v}} T_{a} (\Delta T_a)\dd{A}\\=\int_0^{\pi/2}\left((v_{\max}-v_{\min})\sin\theta+v_{\min}\right)\left(\frac{T_{a,\max}-T_{a,\min}}{2}\cos\theta+\frac{T_{a\max}+T_{a,\min}}{2}\right)\left(\frac{T_{a,\max}-T_{a,\min}}{2}\sin\theta\right)\sin\theta\dd{\theta}
    \\=\frac{1}{48}\left(T_{a,\max}-T_{a,\min}\right)\left(11 T_{a,\max}v_{\max}+5T_{a,\min}v_{\max}+(3\pi-7)T_{a,\max}v_{\min}+3(\pi-3)T_{a,\min}v_{\min}\right)
\end{multline}
As for the RHS of \eqref{eq:xi_1}, we evaluate (dropping integrals in $\phi$ and factors of $r_p$ that cancel)
\begin{multline}
    \overline{\abs{\mathbf{v}}}\overline{T_{a,d}}\left(\overline{\Delta T_a}\right)=\int_0^{\pi/2}\abs{\mathbf{v}}\sin\theta\dd{\theta}\int_0^{\pi/2}T_a\sin\theta\dd{\theta}\int_0^{\pi/2}\Delta T_a\sin\theta\dd{\theta}\\
    =\frac{\pi}{32}(T_{a,\max}-T_{a,\min})(3T_{a,\max}+T_{a,\min})\left(\frac{\pi}{4}(v_{\max}-v_{\min})+v_{\min}\right).
\end{multline}
Thus, the ratio of these is
\begin{equation}
    \frac{8\left(11 T_{a,\max}v_{\max}+5T_{a,\min}v_{\max}+(3\pi-7)T_{a,\max}v_{\min}+3(\pi-3)T_{a,\min}v_{\min}\right)}{3\pi(3T_{a,\max}+T_{a,\min})\left(\pi(v_{\max}-v_{\min})+4v_{\min}\right)}=\frac{8(11+5T+(3\pi-7)v+3(\pi-3)Tv)}{3\pi(3+T)(\pi(1-v)+4v)},
\end{equation}
where $T=T_{a,\min}/T_{a,\max}$ and $v=v_{\min}/v_{\max}.$ If we plot this in Figure \ref{fig:app_xi}, we see that is is very nearly unity over all possible temperature and wind speed ratios. Thus, this term can be neglected for our purposes. For a model where more precision is desired, this effect may need to be considered.
However, we still need to consider the conversion from $\overline{\Delta T_a}$ to $(\overline{T_{a,d}}-\overline{T_{a,n}})$. Taking the integral of $T_{a}$ over the dayside and nightside, we find that for this functional form,
\begin{equation}
    \overline{T_{a,d}}=\frac{3T_{a,\max}+T_{a,\min}}{4},\quad\overline{T_{a,n}}=\frac{T_{a,\max}+3T_{a,\min}}{4}.
\end{equation}
We also see that
\begin{equation}
    \overline{\Delta T_a}=\int_0^{\pi/2}\left(\frac{T_{a,\max}-T_{a,\min}}{2}\sin\theta\right)\sin\theta\dd{\theta}=\frac{\pi}{8}(T_{a,\max}-T_{a,\min})=\frac{\pi}{4}\left(\overline{T_{a,d}}-\overline{T_{a,n}}\right).
\end{equation}
Thus we have $\boxed{\xi\approx\frac{\pi}{4}}.$

\begin{figure}
    \centering
    \includegraphics[width=0.5\linewidth]{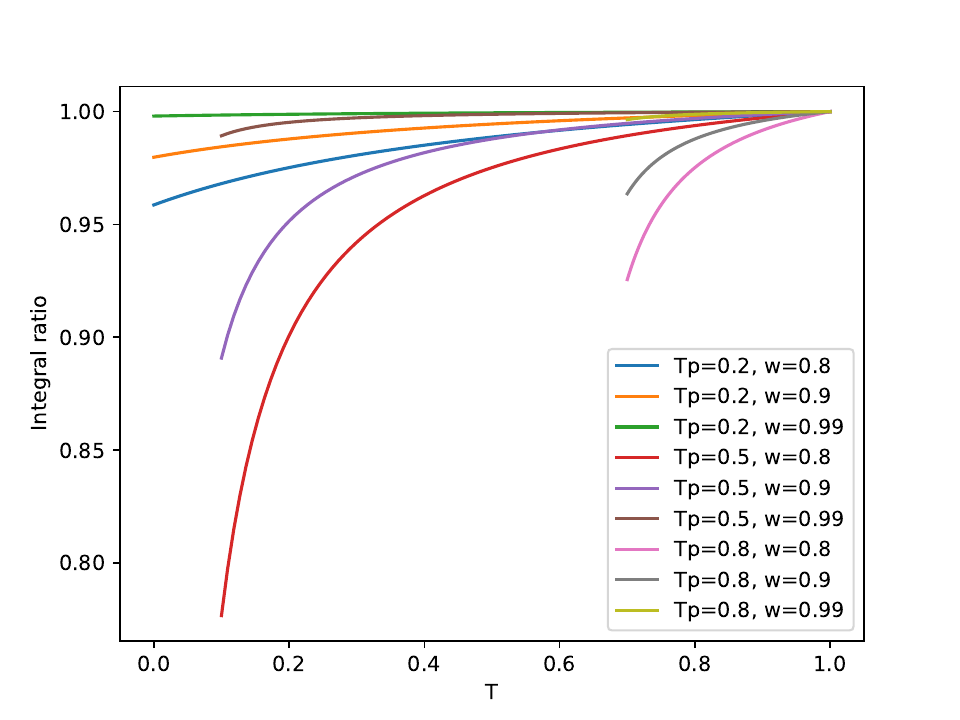}
    \caption{Ratio of the integrals for $\chi$ in Appendix \ref{sec:app_xi}, which is approximately 1 over a large region of parameter space. We stop calculation for values of $T$ less than $2T'-1$ because of division by zero.}
    \label{fig:app_chi}
\end{figure}

We now turn our attention to $\chi,$ which is defined such that
\begin{equation}
    \overline{\abs{\mathbf{w}}(T_{a,d}-T_{s,d})}=\chi\overline{\abs{\mathbf{w}}}\left(\overline{T_{a,d}}-\overline{T_{s,d}}\right),
    \label{eq:app_chi}
\end{equation}
capturing the horizontal weighting in the sensible heat flux term in equations \eqref{eq:localsurf} and \eqref{eq:localatm}. To estimate this value, we assume that variations in $T_{s}$ dominate variations in $T_a$, such that we keep $T_a$ constant and assume a functional form for $T_{s}$ which has maximum $T_{s,\max}$ at the substellar point and $T_{s,t}$ at the terminator:
\begin{equation}
    T_s=(T_{s,\max}-T_{s,t})\cos\theta+T_{s,t}.
\end{equation}
Similarly, we assume $\mathbf{w}$ peaks at the terminator, and we take
\begin{equation}
    \abs{\mathbf{w}}=(w_{\max}-w_{\min})\sin\theta+w_{\min}.
\end{equation}
Thus, the integral for the left side of \eqref{eq:app_chi} is (canceling the integral over $\phi$ and the factors of $r_p$)
\begin{multline}
    \overline{\abs{\mathbf{w}}(T_{a,d}-T_{s,d})}=\int_0^{\pi/2}\left((w_{\max}-w_{\min})\sin\theta+w_{\min}\right)\left(T_a-\left((T_{s,\max}-T_{s,t})\cos\theta+T_{s,t}\right)\right)\sin\theta\dd{\theta}=\\\frac{\pi}{4}(T_a-T_{s,t})(w_{\max}-w_{\min})+\frac{2T_{s,t}w_{\max}-2T_{s,\max}w_{\max}+6T_a w_{\min}-5T_{s,t}w_{\min}-T_{s,\max}w_{\min}}{6}.
\end{multline}
For the right side, we have
\begin{multline}
    \chi\overline{\abs{\mathbf{w}}}\left(\overline{T_{a,d}}-\overline{T_{s,d}}\right)=\int_0^{\pi/2}\abs{\mathbf{w}}\sin\theta\dd{\theta}\left(\int_0^{\pi/2}T_{a}\sin\theta\dd{\theta}-\int_0^{\pi/2}T_{s}\sin\theta\dd{\theta}\right)\\=\left(T_a-\frac{T_{s,\max}-T_{s,t}}{2}\right)\left(\frac{\pi}{4}(w_{\max}-w_{\min})+w_{\min}\right).
\end{multline}
Again, the ratio between these integrals is 
\begin{multline}
    \frac{8(T_{s,\max}-T_{s,t})+6\pi(T_{s,t}-T_a)(w_{\max}-w_{\min})+4(T_{s,\max}+5T_{s,t}-6T_{a})w_{\min}}{3(T_{s,\max}+T_{s,t}-2T_a)(\pi(w_{\max}-w_{\min})+4w_{\min})}\\=\frac{8(1-T)+6\pi(T'-T)(w-1)+4(1+5T-6T')w}{3(1+T-2T')(\pi+4w-\pi w)},
\end{multline}
where $T=T_{s,\min}/T_{s,\max},$ $T'=T_a/T_{s,\max}$, and $w=w_{\min}/w_{\max}$.
We plot this in Figure \ref{fig:app_chi} as a function of $T$ for a few possible values of $w$ and $T'$, where we stop calculation for values of $T$ less than $2T'-1$ because of the shrinking denominator.
We see that, save for areas near a numerical instability, this is also very nearly 1 over most of parameter space. Thus, we neglect its contribution in our analysis and say $\boxed{\chi=1}$. Notably, \citet{wordsworth_atmospheric_2015} found $\chi=1/3,$ but this was due to a lack of normalization by the results of integration over the dayside to find the means. In a more precise analysis, this term may become important, especially when the surface and atmosphere are close in temperature, and so we leave it in our model.

Similar results hold over the nightside of the planet.

\section{Estimation of observational weighting}
\label{sec:app_obs}
To estimate the effect of the larger observational contribution of light coming from the planet close to the substellar point, we assume an intensity functional form similar to those above, where 
\begin{equation}
I(\theta)=(I_{\max}-I_{\min})\cos\theta+I_{\min}.
\end{equation}
This is the functional form that the intensity of a bare rock surface would take, with $I_{\min}=0$. If we let $I_{\min}=I_{\max},$ we have uniform flux, as expected for a planet with uniform heat redistributions. To allow for comparison with K22, we want a conversion factor that interpolates between these regimes, following the assumptions of that work.
We calculate the integrated flux from this assumed profile over the planet's dayside, following \citet{cowan_inverting_2008}:
\begin{equation}
    F_{\mathrm{obs}}=\int_{0}^{2\pi}\int_0^{\pi/2}I(\theta)\cos\theta\sin\theta\dd{\theta}\dd{\phi}=\frac{\pi}{3}\left(2I_{\max}+I_{\min}\right).
\end{equation}
We want to compare this with the result under the assumption of a uniform dayside intensity, so we calculate the mean emission from this parameterization:
\begin{equation}
    \overline{I}=\frac{1}{2\pi}\int_0^{2\pi}\int_0^{\pi/2}I(\theta)\sin\theta\dd{\theta}\dd{\phi}=\frac{I_{\min}+I_{\max}}{2}.
\end{equation}
Using this as the uniform dayside intensity would give us an observed flux of
\begin{equation}
F_{\mathrm{obs,unif}}=\int_{0}^{2\pi}\int_0^{\pi/2}I(\theta)\cos\theta\sin\theta\dd{\theta}\dd{\phi}=\frac{\pi}{2}\left(I_{\max}+I_{\min}\right).
\end{equation}
Thus, the ratio between these two fluxes is
\begin{equation}
    \frac{F_\mathrm{obs}}{F_\mathrm{obs,unif}}=\frac{2(2I_{\max}+I_{\min})}{3(I_{\max}+I_{\min})}.
    \label{eq:app_fluxrat}
\end{equation}
We see that if $I_{\min}=0,$ this ratio becomes $4/3,$ as expected, and if $I_{\max}=I_{\min},$ it is equal to 1. We can put this in terms of $\tilde{T}_2,$ the ratio of the mean dayside to mean nightside temperature as defined in Section \ref{sec:model}, by assuming that the nightside has a uniform intensity of $I_{\min}$, such that we can write 
\begin{equation}
\tilde{T}_2^4=\frac{I_{\min}}{(I_{\max}+I_{\min})/2}.
\end{equation}
Using this, we can put equation (\ref{eq:app_fluxrat}) in terms of $\tilde{T}_2$, which gives us
\begin{equation}
    \frac{F_\mathrm{obs}}{F_\mathrm{obs,unif}}=\frac{4-\tilde{T}_2^4}{3}.
\end{equation}
This reduces to the correct limits in the zero-redistribution and the full-redistribution cases. It is important to note that this is a very rough scaling for the purposes of comparison with other models, and is not very physically motivated. In particular, effects such as the differing contributions of the surface vs. the atmosphere, latitudinal temperature asymmetries due to planetary rotation, and non-uniform nightsides should all be expected to influence this calculation. Further study should be done to better understand the full effect of the viewing geometry on the observed flux from a planet.


\bibliography{references}{}
\bibliographystyle{aasjournal}



\end{document}